\documentclass[twocolumn,tighten,trackchanges]{aastex631}
\PassOptionsToPackage{dvipsnames,svgnames,table}{xcolor}

\usepackage{amsmath}
\usepackage{graphicx}
\usepackage[dvipsnames]{xcolor}
\usepackage{apjfonts}
\usepackage{booktabs}
\usepackage{tablefootnote}

\newcommand{\kms}{\,km\,s$^{-1}$}

\def\lsim{\hbox{\rlap{\raise 0.425ex\hbox{$<$}}\lower 0.65ex\hbox{$\sim$}}}
\def\gsim{\hbox{\rlap{\raise 0.425ex\hbox{$>$}}\lower 0.65ex\hbox{$\sim$}}}

\def\arcdeg{\mbox{$^\circ$}}

\newcommand{\pul}{\text{SN~2022pul}}

\definecolor{maroon}{rgb}{0.760,0.118,0.337}

\newcommand{\LCO}{\affiliation{Las Cumbres Observatory, 6740 Cortona Drive, Suite 102, Goleta, CA 93117-5575, USA}}
\newcommand{\UCSB}{\affiliation{Department of Physics, University of California, Santa Barbara, CA 93106-9530, USA}}

\newcommand{\UCD}{\affiliation{Department of Physics and Astronomy, University of California, Davis, 1 Shields Avenue, Davis, CA 95616-5270, USA}}

\newcommand{\OKC}{\affiliation{Oskar Klein Centre, Department of Physics, Stockholm University, Albanova University Center, SE-106 91, Stockholm, Sweden}}

\newcommand{\UCB}{\affiliation{Department of Astronomy, University of California, Berkeley, CA 94720-3411, USA}}

\newcommand{\STScI}{\affiliation{Space Telescope Science Institute, 3700 San Martin Drive, Baltimore, MD 21218-2410, USA}}

\newcommand{\UA}{\affiliation{Steward Observatory, University of Arizona, 933 North Cherry Avenue, Tucson, AZ 85721-0065, USA}}

\newcommand{\Carnegie}{\affiliation{Observatories of the Carnegie Institute for Science, 813 Santa Barbara Street, Pasadena, CA 91101-1232, USA}}

\newcommand{\IfA}{\affiliation{Institute for Astronomy, University of Hawai`i, 2680 Woodlawn Drive, Honolulu, HI 96822-1839, USA}}
\newcommand{\UCSC}{\affiliation{Department of Astronomy and Astrophysics, University of California, Santa Cruz, CA 95064-1077, USA}}

\newcommand{\Princeton}{\affiliation{Department of Astrophysical Sciences, Princeton University, 4 Ivy Lane, Princeton, NJ 08540-7219, USA}}

\newcommand{\JHU}{\affiliation{Department of Physics and Astronomy, The Johns Hopkins University, 3400 North Charles Street, Baltimore, MD 21218, USA}}

\newcommand{\GeminiNorth}{\affiliation{Gemini Observatory/NSF's NOIRLab, 670 North A`ohoku Place, Hilo, HI 96720-2700, USA}}

\newcommand{\Rutgers}{\affiliation{Department of Physics and Astronomy, Rutgers, the State University of New Jersey,\\136 Frelinghuysen Road, Piscataway, NJ 08854-8019, USA}}
\newcommand{\FSU}{\affiliation{Department of Physics, Florida State University, 77 Chieftan Way, Tallahassee, FL 32306-4350, USA}}
\newcommand{\Melbourne}{\affiliation{School of Physics, The University of Melbourne, Parkville, VIC 3010, Australia}}
\newcommand{\ASTROthreeD}{\affiliation{ARC Centre of Excellence for All Sky Astrophysics in 3 Dimensions (ASTRO 3D)}}

\newcommand{\TAMU}{\affiliation{Department of Physics and Astronomy, Texas A\&M University, 4242 TAMU, College Station, TX 77843, USA}}
\newcommand{\Mitchell}{\affiliation{George P.\ and Cynthia Woods Mitchell Institute for Fundamental Physics \& Astronomy, College Station, TX 77843, USA}}

\newcommand{\ICE}{\affiliation{Institute of Space Sciences (ICE, CSIC), Campus UAB, Carrer de Can Magrans, s/n, E-08193 Barcelona, Spain}}
\newcommand{\IEEC}{\affiliation{Institut d’Estudis Espacials de Catalunya (IEEC), E-08034 Barcelona, Spain}}

\newcommand{\MSUPA}{\affiliation{Department of Physics and Astronomy, Michigan State University, East Lansing, MI 48824, USA}}
\newcommand{\MSUCMSE}{\affiliation{Department of Computational Mathematics, Science, and Engineering, Michigan State University, East Lansing, MI 48824, USA}}
\newcommand{\UPitt}{\affiliation{Department of Physics and Astronomy and Pittsburgh Particle Physics, Astrophysics and Cosmology Center (PITT PACC), University of Pittsburgh, 3941 O’Hara Street, Pittsburgh, PA 15260, USA}}

\newcommand{\USzeged}{\affiliation{Department of Experimental Physics, Institute of Physics, University of Szeged, D\'om t\'er 9, 6720 Szeged, Hungary}}
\newcommand{\ELKHSZTE}{\affiliation{ELKH-SZTE Stellar Astrophysics Research Group, Szegedi \'ut, Kt. 766, 6500 Baja, Hungary}}
\newcommand{\Konkoly}{\affiliation{Konkoly Observatory, Research Centre for Astronomy and Earth Sciences (CSFK), MTA Center of Excellence, Konkoly-Thege Mikl\'os \'ut 15-17, 1121 Budapest, Hungary}}

\newcommand{\AMNH}{\affiliation{Department of Astrophysics, American Museum of Natural History, Central Park West and 79th Street, New York, NY 10024-5192, USA}}
\newcommand{\UDublin}{\affiliation{School of Physics, Trinity College Dublin, The University of Dublin, Dublin 2, Ireland}}
\newcommand{\VTech}{\affiliation{Department of Physics, Virginia Tech, Blacksburg, VA 24061, USA}}
\newcommand{\USheffield}{\affiliation{Department of Physics and Astronomy, University of Sheffield, Hicks Building, Hounsfield Road, Sheffield S3 7RH, U.K.}}
\newcommand{\Aarhus}{\affiliation{Department of Physics and Astronomy, Aarhus University, Ny Munkegade 120, DK-8000 Aarhus C, Denmark}}

\newcommand{\ICG}{\affiliation{Institute of Cosmology \& Gravitation, University of Portsmouth, Dennis Sciama Building, Burnaby Road, Portsmouth PO1 3FX, UK}}

\newcommand{\Liverpool}{\affiliation{Astrophysics Research Institute, Liverpool John Moores University, Liverpool,  L3 5RF, UK}}
\newcommand{\MaxPlanck}{\affiliation{Max-Planck Institute for Astrophysics, Garching, Germany}}
\newcommand{\GSI}{\affiliation{GSI Helmholtzzentrum f\"ur Schwerionenforschung, Planckstra\ss{}e 1, 64291 Darmstadt, Germany}}
\newcommand{\KyotoU}{\affiliation{Department of Astronomy, Kyoto University, Kitashirakawa-Oiwake-cho, Sakyo-ku, Kyoto, 606-8502. Japan}}
\newcommand{\OSU}{\affiliation{Center for Cosmology and Astroparticle Physics, The Ohio State University, 191 West Woodruff Ave, Columbus, OH 43215, USA}}
\newcommand{\LasCampanas}{\affiliation{Carnegie Observatories, Las Campanas Observatory, Casilla 601, La Serena, Chile}}
\newcommand{\NotreDame}{\affiliation{Department of Physics and Astronomy, University of Notre Dame, Notre Dame, IN 46556, USA}}

\newcommand{\AixMarseille}{\affiliation{Aix Marseille Univ, CNRS, CNES, LAM, Marseille, France}}
\newcommand{\IAP}{\affiliation{Institut d'Astrophysique de Paris, CNRS-Sorbonne Universit\'{e}, 98 bis boulevard Arago, 75014, Paris, France}}
\newcommand{\Thailand}{\affiliation{National Astronomical Research Institute of Thailand, 260 Moo 4, Donkaew, Maerim, Chiang Mai 50180, Thailand}}
\newcommand{\PUC}{\affiliation{Instituto de Astrof\'{i}sica, Facultad de F\'{i}sica, Pontificia Universidad Cat\'{o}lica de Chile, Av. Vicu\~{n}a Mackenna 4860, Santiago, Chile}}
\newcommand{\MIA}{\affiliation{Millennium Institute of Astrophysics, Nuncio Monse\~{n}or S\'{o}tero Sanz 100, Providencia, Santiago, Chile}}
\newcommand{\TAPIR}{\affiliation{TAPIR, Walter Burke Institute for Theoretical Physics, 350-17, Caltech, Pasadena, CA 91125, USA}}
\newcommand{\UTexas}{\affiliation{Department of Astronomy, University of Texas at Austin, Austin, TX, USA}}
\newcommand{\VVS}{\affiliation{Vereniging Voor Sterrenkunde (VVS), Oostmeers 122 C, 8000 Brugge, Belgium}}
\newcommand{\AAVSO}{\affiliation{AAVSO, 185 Alewife Brook Parkway, Suite 410, Cambridge, MA 02138, USA}}
\newcommand{\GEOS}{\affiliation{Groupe Europ\'{e}en d'Observations Stellaires (GEOS), 23 Parc de Levesville, 28300 Bailleau l'Ev\^{e}que, France}}
\newcommand{\BAV}{\affiliation{Bundesdeutsche Arbeitsgemeinschaft für Ver\"{a}nderliche Sterne (BAV), Munsterdamm 90, 12169 Berlin, Germany}}
\newcommand{\LPNHE}{\affiliation{LPNHE, (CNRS/IN2P3, Sorbonne Universit\'{e}, Universit\'{e} Paris Cit\'{e}), Laboratoire de Physique Nucl\'{e}aire et de Hautes \'{E}nergies, 75005, Paris, France}}
\newcommand{\ING}{\affiliation{Isaac Newton Group (ING), Apt. de correos 321, E-38700, Santa Cruz de La Palma, Canary Islands, Spain}}
\newcommand{\ELTE}{\affiliation{ELTE E\"otv\"os Lor\'and University, Institute of Physics and Astronomy, P\'azm\'any P\'eter s\'et\'any 1/A, Budapest, 1117 Hungary}}

\shorttitle{Unusual Signatures in SN~2022pul}
\shortauthors{Siebert et al.}

\submitjournal{\textit{The Astrophysical Journal}}

\graphicspath{{./}{figures/}}

\begin{document}

\title{Ground-based and JWST Observations of SN~2022pul: \\ 
I. Unusual Signatures of Carbon, Oxygen, and Circumstellar Interaction in a Peculiar Type Ia Supernova}

\correspondingauthor{Matthew~R.~Siebert}
\email{msiebert@stsci.edu}

%% first 3 will get shuffled for papers I, II, III
\author[0000-0003-2445-3891]{Matthew~R.~Siebert}
\STScI
\author[0000-0003-3108-1328]{Lindsey A.\ Kwok}
\Rutgers
\author[0000-0001-5975-290X]{Joel Johansson}
\OKC
%% all the rest should be consistent across all three papers
\author[0000-0001-8738-6011]{Saurabh W.\ Jha}
\Rutgers
\author[0000-0002-9388-2932]{St\'{e}phane Blondin}
\AixMarseille
\author[0000-0003-0599-8407]{Luc Dessart}
\IAP
\author[0000-0002-2445-5275]{Ryan J.\ Foley}
\UCSC
\author[0000-0001-5094-8017]{D.\ John Hillier}
\UPitt
\author[0000-0003-2037-4619]{Conor Larison}
\Rutgers
\author[0000-0003-3308-2420]{R\"{u}diger Pakmor}
\MaxPlanck
\author[0000-0001-7380-3144]{Tea Temim}
\Princeton
\author[0000-0003-0123-0062]{Jennifer E.\ Andrews}
\GeminiNorth
\author[0000-0002-4449-9152]{Katie Auchettl}
\Melbourne
\UCSC
\author[0000-0003-3494-343X]{Carles Badenes}
\UPitt
\author[0000-0003-4769-4794]{Barnabas Barna}
\USzeged
\author[0000-0002-4924-444X]{K.\ Azalee Bostroem}
\thanks{LSSTC Catalyst Fellow}
\UA
\author[0000-0002-8092-2077]{Max J.\ Brenner Newman}
\Rutgers
\author[0000-0001-5955-2502]{Thomas G.\ Brink}
\UCB
\author[0000-0003-0416-9818]{Mar\'{i}a Jos\'{e} Bustamante-Rosell}
\UCSC
\author[0000-0002-9830-3880]{Yssavo Camacho-Neves}
\Rutgers
\author[0000-0003-3068-4258]{Alejandro Clocchiatti}
\PUC
\MIA
\author[0000-0003-4263-2228]{David~A.~Coulter}
\UCSC
\author[0000-0002-5680-4660]{Kyle W.\ Davis}
\UCSC
\author[0000-0001-8857-9843]{Maxime Deckers}
\UDublin
\author[0000-0001-9494-179X]{Georgios Dimitriadis}
\UDublin
\author[0000-0002-7937-6371]{Yize Dong}
\UCD
\author[0000-0003-4914-5625]{Joseph Farah}
\LCO
\UCSB
\author[0000-0003-3460-0103]{Alexei V.\ Filippenko}
\UCB
\author[0000-0003-2024-2819]{Andreas Fl\"ors}
\GSI
\author[0000-0003-2238-1572]{Ori D.\ Fox}
\STScI
\author[0000-0003-4069-2817]{Peter Garnavich}
\NotreDame
\author[0000-0003-0209-9246]{Estefania Padilla Gonzalez}
\LCO
\UCSB
%\author[0000-0002-9154-3136]{Melissa L.\ Graham} % opted out
%\DiRAC
\author[0000-0002-4391-6137]{Or Graur}
\ICG
\AMNH
\author[0000-0003-0125-8700]{Franz-Josef Hambsch}
\VVS
\AAVSO
\GEOS
\BAV
\author[0000-0002-0832-2974]{Griffin Hosseinzadeh}
\UA
\author[0000-0003-4253-656X]{D.\ Andrew Howell}
\LCO
\UCSB
\author[0000-0002-8816-6800]{John P.\ Hughes}
\Rutgers
%\author[0000-0002-7612-0469]{Sarah Kendrew}     % opted out
%\ESASTScI
\author[0000-0002-0479-7235]{Wolfgang E.\ Kerzendorf}
\MSUPA
\MSUCMSE
\author[0009-0004-3242-282X]{Xavier K.\ Le Saux}
\UCSC
\author[0000-0003-2611-7269]{Keiichi Maeda}
\KyotoU
\author[0000-0002-9770-3508]{Kate Maguire}
\UDublin
\author[0000-0001-5807-7893]{Curtis McCully}
\LCO
\UCSB
\author[0009-0004-0322-6299]{Cassidy Mihalenko} 
\Melbourne
\ASTROthreeD
\author[0000-0001-9570-0584]{Megan Newsome}
\LCO
\UCSB
\author[0000-0003-3615-9593]{John T.\ O'Brien}
\MSUPA
\author[0000-0002-0744-0047]{Jeniveve Pearson}
\UA
\author[0000-0002-7472-1279]{Craig Pellegrino}
\LCO
\UCSB
\author[0000-0002-2361-7201]{Justin D.\ R.\ Pierel}
\STScI
\author[0000-0002-1633-6495]{Abigail Polin}
\Carnegie
\TAPIR
\author[0000-0002-4410-5387]{Armin Rest}
\STScI
\JHU
\author[0000-0002-7559-315X]{C\'{e}sar Rojas-Bravo}
\UCSC
\author[0000-0003-4102-380X]{David J.\ Sand}
\UA
\author[0009-0002-5096-1689]{Michaela Schwab}
\Rutgers
\author[0000-0002-9301-5302]{Melissa Shahbandeh}
\STScI
\author[0000-0002-4022-1874]{Manisha Shrestha}
\UA
\author[0000-0001-5510-2424]{Nathan Smith}
\UA
\author[0000-0002-7756-4440]{Louis-Gregory Strolger}
\STScI
\author[0000-0003-4610-1117]{Tam\'as Szalai}
\USzeged
\ELKHSZTE
\author[0000-0002-5748-4558]{Kirsty Taggart}
\UCSC
\author[0000-0003-0794-5982]{Giacomo Terreran}
\LCO
\UCSB
\author[0000-0001-9834-3439]{Jacco H.\ Terwel}
\UDublin
\ING
\author[0000-0002-1481-4676]{Samaporn Tinyanont}
\Thailand
\author[0000-0001-8818-0795]{Stefano Valenti}
\UCD
\author[0000-0001-8764-7832]{J\'{o}zsef Vink\'{o}}
\Konkoly
\USzeged
\ELTE
\UTexas
\author[0000-0003-1349-6538]{J.\ Craig Wheeler}
\UTexas
\author[0000-0002-6535-8500]{Yi Yang}
\UCB
\author[0000-0002-2636-6508]{WeiKang Zheng}
\UCB
%
% commented out folks from the Ashall team who haven't yet responded to the
% email; let's make sure everyone has ample opportunity to opt back in!
%
\author[0000-0002-5221-7557]{Chris Ashall}
\VTech
%\author[0000-0001-5393-1608]{E. Baron}
%\UOklahoma
%\Hamburg
%\author[0000-0003-4625-6629]{Chris R.\ Burns}
%\Carnegie
\author[0000-0002-7566-6080]{James M.\ DerKacy}
\VTech
%\author[0000-0001-5888-2542]{Tyco Mera Evans}
%\FSU
%\author[0000-0002-5253-3584]{Alec Fisher}
%\FSU
\author[0000-0002-1296-6887]{Llu\'is Galbany}
\ICE
\IEEC
\author[0000-0002-4338-6586]{Peter Hoeflich}
\FSU
\author[0000-0003-1039-2928]{Eric Hsiao}
\FSU
\author[0000-0001-6069-1139]{Thomas de Jaeger}
\LPNHE
%\author[0000-0001-6209-838X]{Emir Karamehmetoglu}
%\Aarhus
%\author[0000-0002-6650-694X]{Kevin Krisciunas}
%\TAMU
%\Mitchell
%\author[0000-0001-8367-7591]{Sahana Kumar}
%\FSU
\author[0000-0002-3900-1452]{Jing Lu}
\MSUPA
\author[0000-0003-0733-7215]{Justyn Maund}
\USheffield
%\author[0000-0001-6876-8284]{Paolo A.\ Mazzali}
%\Liverpool
%\MaxPlanck
\author[0000-0001-7186-105X]{Kyle Medler}
\Liverpool
\author[0000-0003-2535-3091]{Nidia Morrell}
\LasCampanas
%\author[0000-0003-2734-0796]{Mark. M.\ Phillips}
%\LasCampanas
\author[0000-0003-4631-1149]{Benjamin J.\ Shappee} 
\IfA
\author[0000-0002-5571-1833]{Maximilian Stritzinger}
\Aarhus
\author[0000-0002-8102-181X]{Nicholas Suntzeff}
\TAMU
\Mitchell
%\author[0000-0002-0036-9292]{Charles Telesco}
%\UFlorida
\author[0000-0002-2471-8442]{Michael Tucker}
\thanks{CCAPP Fellow}
\OSU
\author[0000-0001-7092-9374]{Lifan Wang}
\TAMU
\Mitchell

\begin{abstract}
Nebular-phase observations of peculiar Type Ia supernovae (SNe~Ia) provide important constraints on progenitor scenarios and explosion dynamics for both these rare SNe and the more common, cosmologically useful SNe~Ia. We present observations from an extensive ground-based and space-based follow-up campaign to characterize \pul, a ``super-Chandrasekhar'' mass SN~Ia (alternatively ``03fg-like'' SN), from before peak brightness to well into the nebular phase across optical to mid-infrared (MIR) wavelengths. The early rise of the light curve is atypical, exhibiting two distinct components, consistent with SN~Ia ejecta interacting with dense carbon-oxygen rich circumstellar material (CSM). In the optical, \pul\ is most similar to SN~2012dn, having a low estimated peak luminosity ($M_{B}=-18.9$~mag) and high photospheric velocity relative to other 03fg-like SNe. In the nebular phase, \pul\ adds to the increasing diversity of the 03fg-like subclass. From 168 to 336 days after peak $B$-band brightness, \pul\ exhibits asymmetric and narrow emission from [\ion{O}{1}] $\lambda\lambda 6300$, 6364 (${\rm FWHM} \approx 2{,}000$\kms), strong, broad emission from [\ion{Ca}{2}] $\lambda\lambda 7291$, 7323 (${\rm FWHM} \approx 7{,}300$\kms), and a rapid \ion{Fe}{3} to \ion{Fe}{2} ionization change. Finally, we present the first-ever optical-to-mid-infrared (MIR) nebular spectrum of an 03fg-like SN~Ia using data from \textit{JWST}. In the MIR, strong lines of neon and argon, weak emission from stable nickel, and strong thermal dust emission (with $T \approx 500$~K), combined with prominent [\ion{O}{1}] in the optical, suggest that \pul\ was produced by a white dwarf merger within carbon/oxygen-rich CSM.

\end{abstract}

\keywords{Supernovae Type Ia --- white dwarf --- thermonuclear explosions --- ejecta mass }

\section{Introduction}\label{s:intro}

Type Ia supernovae (SNe~Ia) are standardizable candles via the strong and well-studied correlations of their luminosities with their light-curve shapes and colors \citep{Phillips93, Tripp98}. Because of this, and their intrinsically high luminosities, they are excellent cosmological distance indicators and have been used to show that the expansion of the Universe is accelerating \citep{Riess98:lambda, Perlmutter99}.

Despite their extensive cosmological use, no consensus exists as to what progenitor systems and explosion mechanisms can explain the majority of ``normal'' SNe~Ia --- those used for cosmological analyses. The debate is primarily centered around two broad classes of binary systems, the single-degenerate (SD) and double-degenerate (DD) channels. In an SD progenitor system, one possible pathway is when a carbon-oxygen (C/O) white dwarf (WD) reaches near the ``Chandrasekhar mass'' ($M_{\rm Ch} \approx 1.4\ M_{\odot}$, \citealt{Chandrasekhar31}) by accreting from a nondegenerate star such as a giant, subgiant, main-sequence, or subdwarf star \citep{Whelan73}, achieving the central densities and temperatures needed to ignite runaway carbon burning \citep{Seitenzahl13:3d,Branch_Wheeler17, Kerzendorf17}. In a DD progenitor system, the explosion can occur via the merger of two WDs \citep{IbenTotukov84, Webbink84}. In the latter system, the explosion may occur ``violently'' during the merger \citep{Pakmor10,Pakmor12} or be significantly delayed, resulting in an explosion long after disruption of the secondary WD \citep{Raskin13,Dan14}. \citet{Schwab16} showed that merger remnants in excess of $M_{\rm Ch}$ may not achieve the densities necessary for an explosion and instead will collapse to a neutron star, and \citet{Dan14} found that only the most massive mergers ($M_{\rm tot} \geq 2.1 \ M_{\odot}$) would lead to a detonation. 

An alternative explosion mechanism in either the SD or DD progenitor systems is a ``double detonation'' where a surface He layer explosively burns, causing a second explosion in the interior of the WD \citep{Woosley11, Nomoto18, Polin19}. This provides a potential pathway for the detonation of sub-$M_{\rm Ch}$ WDs. Some studies have found that double detonations of sub-$M_{\rm Ch}$ WDs in DD systems with little to no He on their surface can explain the observed properties of normal SNe~Ia \citep{Shen14, Townsley19, Leung20}. 

Several WD explosions, including subclasses of SNe~Ia, do not follow the tight relationships between luminosity, light-curve shape, and color \citep[for reviews, see][]{Taubenberger17,Jha+19,Ashall21}. The unique characteristics of these extreme thermonuclear SN classes can inform progenitor models of normal SNe~Ia, and potential sources of contamination in samples used to measure cosmological distances.

One particularly important peculiar class is the rare, carbon-strong, high-luminosity or ``03fg-like'' SNe that occupy the brightest end of SNe~Ia \citep{Howell06,Hicken07,Yamanaka09,Tanaka10,Scalzo10,Silverman11:09dc}. This class is differentiated from normal SNe~Ia by their relatively high luminosities (ranging from $M_{B} = -19.1$ to $-20.3$~mag), strong and persistent \ion{C}{2} absorption at early phases, low photospheric velocities \citep{Howell06}, rapidly evolving early-time light-curve bumps \citep{Jiang+21_SN20hvf, Srivastav23,Dimitriadis23}, delayed times of $i$-band peak brightness relative to $B$-band peak brightness \citep{Ashall20}, faster optical fading at late times coincident with excess emission in the near-infrared (NIR), and low-ionization-state nebular spectra including strong [\ion{Ca}{2}] emission  (and in some cases [\ion{O}{1}]; \citealt{Kromer16, Dimitriadis23, Siebert23a}). 

If powered by the radioactive decay of $^{56}$Ni alone, the optical light curves of these SNe suggest an explosion that produced an ejecta mass larger than $M_{\rm Ch}$ \citep{Howell06, Scalzo10, Dimitriadis22}. Like normal SNe~Ia, both SD and DD channels have been proposed as potential origins of these events. In the SD scenario, a possible interpretation was the explosion of a WD that is supported by rapid rotation \citep{Yoon05}. It is possible that the higher binding energy of a massive WD would result in lower ejecta velocities \citep{Howell06}, but this has not yet been realized in simulations. In particular, the high ejecta velocities produced by models of rapidly-rotating massive WD progenitors \citep{maeda09b,Hachinger12:sc,Fink18} are in conflict with the observed absorption velocities of 03fg-like SNe~Ia at early times. In the DD scenario, a total ejecta mass greater than $M_{\rm Ch}$ can be naturally explained by the merger of two sub-$M_{\rm Ch}$ C/O WDs ($M \gtrsim 0.7$~M$_{\sun}$ for each). In this scenario, the tidal disruption of the secondary WD could lead to the the formation of a disk or shell of C/O-rich material \citep{Raskin13}. 

The presence of dense pre-existing or newly ejected material in the vicinity of the explosion can explain several observed properties of 03fg-like SNe~Ia. The rapid early light-curve bumps seen for three events \citep{Jiang+21_SN20hvf, Srivastav23,Dimitriadis23} may result from the initial interaction with this circumstellar material (CSM). Alternatively, prolonged interaction can increase the peak luminosity beyond a typical SN~Ia \citep{Hicken07, Hachinger12:sc,Taubenberger13:sc, Noebauer16}. The spectra of the 03fg-like SN~2020esm obtained only days after explosion displayed only C and O features, suggesting a predominantly C/O atmosphere in the outermost layers of the ejecta, consistent with swept-up C/O-rich CSM \citep{Dimitriadis22}.  The deceleration of the ejecta by C/O-rich CSM helps to explain the low ejecta velocities at peak and the persistent \ion{C}{2} lines. Additionally, the rapid optical fading \citep{Taubenberger19,Dimitriadis22} coincident with a NIR excess at late times was observed for SN 2012dn \citep{Yamanaka16,Nagao17,Nagao18}; it is potentially explained by the formation of dust in the SN ejecta, but the origin of the dust is still controversial. 

Multiple avenues have been proposed to explain the presence of this CSM. In a DD scenario, H/He-poor (and C/O-rich) CSM could be produced via the dynamical interaction of two WDs \citep{Raskin13}. Tidal stripping of the secondary star during a merger could leave unburned, H/He-poor material in the vicinity of the progenitor system prior to explosion. This scenario has been favored in several studies of 03fg-like SNe~Ia \citep{Dimitriadis22,Srivastav23, Siebert23a}. 
Alternatively, the CSM may be explained by the
C-rich envelope of an asymptotic giant branch (AGB) star. The progenitor system proposed in this case (the ``core-degenerate" scenario) instead requires the explosion of a near-$M_{\rm Ch}$ C/O WD \citep{Kashi_Soker11,Hsiao20,Lu21}.

Neither of these scenarios can provide a perfect match to all observed properties of 03fg-like SNe. In particular, a merger of two WDs is an inherently asymmetric scenario and is expected to produce large levels of continuum polarization \citep{Bulla16}, in contrast to the low continuum polarization observed for both SN~2007if and SN~2009dc, suggesting more spherical explosions \citep{Tanaka10}. An SN~Ia explosion within the envelope of an AGB star should exhibit a prolonged interaction phase producing excess X-ray and ultraviolet (UV) luminosity at later times \citep{Hsiao20}, which is not seen. Additionally, we have not yet observed narrow H/He emission which would be expected from the interaction of the ejecta with the envelope of the AGB star \citep{Kashi_Soker11,Hsiao20, Lu21}. However, this may be possible if the H/He was lost in a superwind phase prior to the explosion. Another distinction between these models is that a Chandrasekhar-mass WD would be expected to undergo high-density burning and produce more stable nickel \citep{Hoflich96, Iwamoto99, Seitenzahl13:3d}. The presence of [\ion{Ni}{2}] is especially difficult to constrain in optical nebular spectra of 03fg-like SNe because this feature is heavily blended with the often very strong emission from [\ion{Ca}{2}]. 

One important clue to the origin of these enigmatic SNe is the presence of [\ion{O}{1}] emission in several of their nebular spectra. Understanding oxygen emission in WD SNe is difficult since it requires knowing whether the origin of the emitting material is from an unburned part of the WD or produced during carbon burning. Currently, late-time [\ion{O}{1}] emission has only been observed in rare subclasses of SNe~Ia. Specifically, narrow [\ion{O}{1}] emission has been observed in the low-luminosity SN 2002es-like \citep{Ganeshalingam12} SNe~2010lp and iPTF14atg \citep{Kromer16}. This was attributed to the violent merger of two WDs since the presence of oxygen at low velocities is a natural prediction of these models \citep{Pakmor12, Mazzali22}. [\ion{O}{1}] emission has also been observed \citep{Prentice22} in low-luminosity ``Ca-rich SNe" \citep{Filippenko03:carich}, which may also originate from DD scenarios \citep{Shen19,Jacobson-Galan19, Zenati+23_carich}.

Finally, 03fg-like SNe are a diverse subclass. These SNe cover a large range of peak luminosities which are not correlated with their light-curve shapes \citep{Ashall21}. Their late-time light curves experience faster decline rates that begin at various epochs, and emission features in their nebular spectra vary greatly in their strength, width, symmetry, and velocity offset \citep{Siebert23a}. A viable progenitor scenario should be able to explain this diversity. 

Here, we present results from an extensive ground- and space-based follow-up campaign to observe the 03fg-like SN~Ia~2022pul. This SN was discovered on 28 July 2022 (UTC dates are used throughout this paper) by the All-Sky Automated Survey for Supernovae (ASAS-SN; \citealt{Shappee14}). The host galaxy, NGC~4415, has a redshift of $z = 0.00301$ \citep{Francois19} and a distance of 16~Mpc (see \autoref{s:dist}). Like other 03fg-like SNe, this SN has a large spatial offset from its host galaxy, possibly indicating that it could have originated in a low-metallicity environment \citep{Khan11}. SN~2022pul was quickly classified as an SN~Ia by the Spectroscopic Classification of Astronomical Transients (SCAT) Survey \citep{Tucker22b} on 1 Aug.~2022. 

This paper is organized as follows.
In \autoref{s:obs} we present our photometric and spectroscopic time series of \pul. We then show the first continuous optical to mid-infrared (MIR) spectrum of an 03fg-like SN~Ia achieved through observations from \textit{JWST} (Cycle 1 GO-2072; PI S.~W.~Jha). In \autoref{s:anal} we compare the properties of \pul\ to those of other peculiar SNe~Ia and compare to models. This SN adds to the growing diversity of 03fg-like SNe~Ia. Our early multiband light curves exhibit an initial rise that deviates from a smooth evolution, indicative of CSM interaction. Our late-time optical spectroscopic sequence reveals a quickly changing ionization state, strong and broad emission from [\ion{Ca}{2}], and unprecedented narrow emission from [\ion{O}{1}]. Asymmetric nebular emission profiles are observed in both the optical and NIR. In the MIR, we observe low-ionization nebular emission lines, and a strong thermal ($T\approx500$~K) dust continuum. Our data are consistent with an asymmetric explosion that did not occur within a pristine environment.  These results are discussed in detail in Paper II \citep{Kwok23b} and Paper III \citep{Johansson23}. 

In \autoref{s:disc}, we discuss all of our observations and consider their implications for the progenitor system of \pul\ and the subclass of 03fg-like SNe~Ia. We conclude and summarize the results of our observations in \autoref{s:conc}.

\section{Observations \& Data Reduction}\label{s:obs}

\subsection{Photometry}
\begin{figure*}[htb!]
    \centering
    \includegraphics[width=\textwidth]{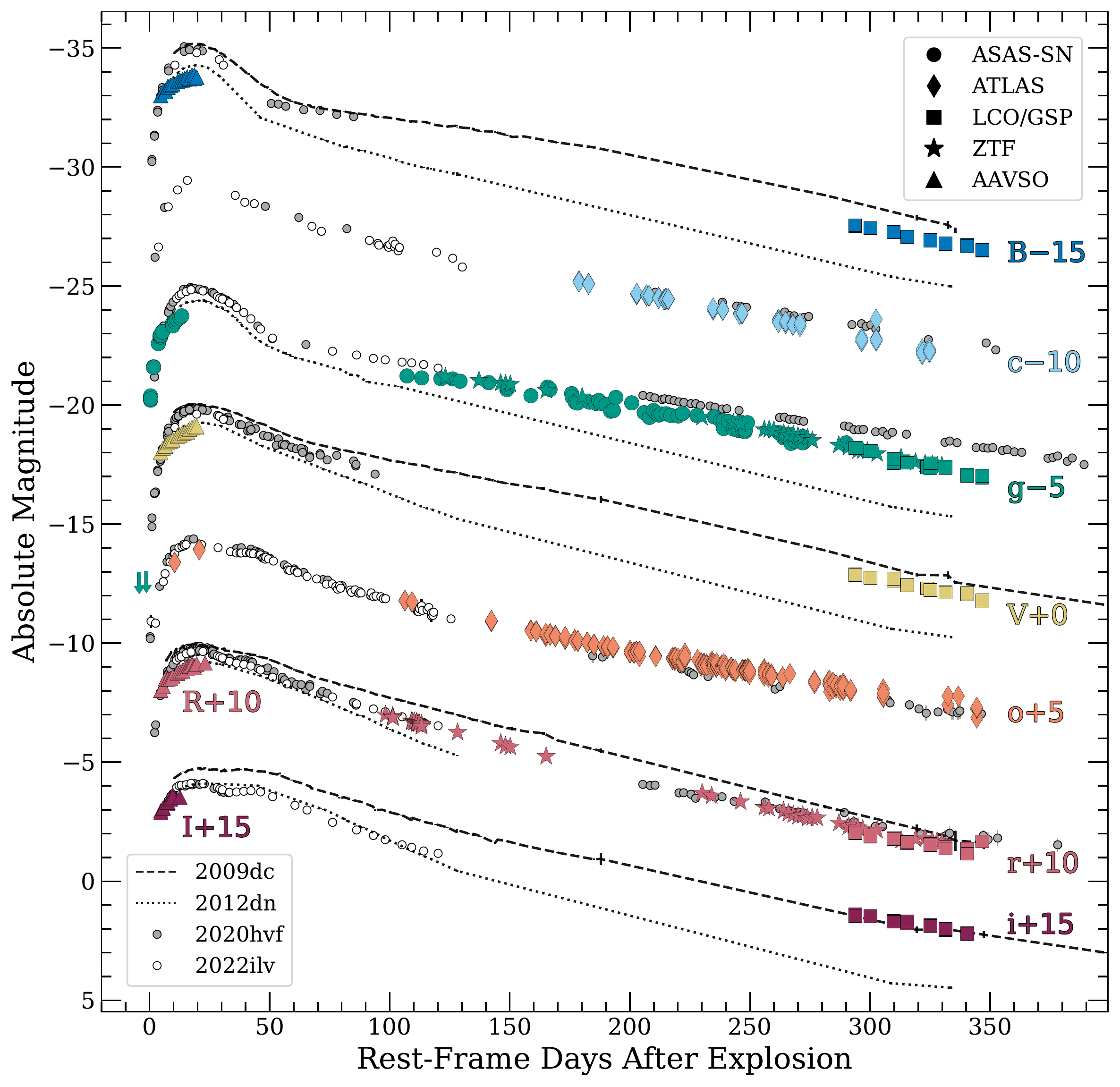}
    \caption{Multiband light curve of \pul\ in rest-frame days relative to $t_{\text{exp}}= 59786.3$ MJD (colored points) plotted with offsets shown on the right-hand side for visual clarity. The green downward-pointing arrows are upper limits from ASAS-SN. For comparison we show multiband light curves of other 03fg-like SNe 2009dc (\citealt{Taubenberger11}, black-dashed line), 2012dn (\citealt{Taubenberger19}, black-dotted line), 2020hvf (\citealt{Jiang+21_SN20hvf}, gray points), and 2022ilv (\citealt{Srivastav23}, white points). \pul\ was not observable shortly after peak brightness, limiting our ability to measure the light-curve shape. We estimate peak $B$-band brightness of $-18.9$~mag. All SN light curves were corrected for MW and host-galaxy extinction when necessary. 
    \label{fig:lightcurve}}
\end{figure*}

We first gathered photometric observations of \pul\ through our own collaborations. This includes data obtained with the Sinistro cameras of the Las Cumbres Observatory (LCO; \citealt{Brown13}) network of 1~m telescopes through the Global Supernova Project (GSP). The Sinistro images were processed with a dedicated python/pyraf pipeline \citep{Valenti16}\footnote{\url{https://github.com/LCOGT/lcogtsnpipe}}.

We then obtained additional photometry from public databases. These sources include ASAS-SN \citep{Shappee14}, forced photometry from the Asteroid Terrestrial impact Last Alert System (ATLAS; \citealt{Tonry+18_ATLAS, Smith20}) in the cyan and orange bands, the Zwicky Transient Facility (ZTF; \citealt{Bellm19}), and the American Association of Variable Star Observers (AAVSO; Observations from the AAVSO International Database\footnote{\url{https://www.aavso.org}}). ZTF photometry was obtained with the wide-field camera on the 1.2~m Samuel Oschin (P48) Telescope in $g$ and $r$ filters. AAVSO photometry was obtained by the Remote Observatory Atacama Desert (ROAD), which consists of a 0.4~m $f$/6.8 Optical Dall-Kirkham (ODK) telescope with an FLI ML16803 CCD and Astrodon photometric $BVRI$ filters.

Our photometry is presented in \autoref{fig:lightcurve}. Our data range from $\sim 0.4$ to $350$ days after explosion. Unfortunately, around the time of peak brightness, \pul\ was not observable, and this limits our ability to characterize its light-curve shape. All SN light curves were corrected for extinction (see \autoref{s:dist}). AAVSO observations from the international amateur astronomical community provided essential additional data, covering the rise with 4 additional bands.

By fitting a parabola to the ASAS-SN \textit{g}-band rise points of the light curve and solving for the zero-point of the flux, we find $t_\text{explosion} =$ MJD 59786.3 (2022 July 26.3 UTC). The most constraining ASAS-SN \textit{g}-band nondetection was at 59785.00 MJD at a depth of 17.98 mag, and the first detection (ASAS-SN \textit{g}-band) was at 59786.73 MJD. Using our $B$-band light curve, we define peak brightness as $t_{B,{\rm peak}} = t_{\rm explosion}+22$ days. Unless otherwise noted, throughout this work phases are defined relative to $t_{B,{\rm peak}}$.

\subsection{Extinction, Distance, and Luminosity}\label{s:dist}

\pul\ is located in the outskirts of its elliptical host galaxy NGC 4415 in the Virgo cluster and its offset is 2.207 arcminutes (projected distance of 9 kpc). The SDSS $r$-band Petrosian radius of NGC 4415 is 2.3 kpc \citep{SDSS_dr6}, so we assume that the extinction by dust from the host galaxy is minimal and the main source of extinction will come from the local environment of the SN itself. Therefore, we correct all spectra and photometry for Milky Way (MW) extinction of $E(B-V)_{\rm MW} = 0.008$~mag \citep{schlafly_measuring_2011}, but we do not correct for extinction from the host galaxy. We use the Python \texttt{dust-extinction} package \citep[v.~1.1;][]{karl_gordon_2022_6397654}. For our \textit{JWST} spectroscopy, we deredden the NIRSpec spectrum blueward of 1.0~$\mu$m using the F19 model from \citet{Fitzpatrick2019} and the NIRSpec spectrum redward of 1.0~$\mu$m as well as the MIRI spectrum with the G21\_MWavg model from \citet{Gordon2021}.

The distance to NGC 4415 is somewhat uncertain; it is projected close to M49 and thus likely part of the Virgo B subcluster. Using the surface brightness fluctuation (SBF) method, \citet{Jerjen2004} find a distance of $14.9 \pm 1.0$ Mpc, whereas the fundamental plane (FP) gives $17.1 \pm 2.0$ Mpc \citep[$\mu = 31.17 \pm 0.25$~mag;][]{Tully2013}. Combining these with a conservative uncertainty estimate, we adopt a distance of $16 \pm 2$ Mpc (distance modulus $\mu = 31.02 \pm 0.27$ mag) for \pul.

Given the atypical behavior of the early-time light curve, and the lack of data immediately after peak brightness, it is difficult to obtain a precise peak brightness and light-curve shape for \pul. Using the distance above, and assuming that our latest early-phase photometry was near peak brightness, we estimate \pul\ had $M_{B,{\rm peak}} = -18.9$ mag. This would make \pul\ the lowest-luminosity 03fg-like SN~Ia (the previous lowest being ASASSN-15hy, $M_{B,{\rm peak}} = -19.14$ mag; \citealt{Lu21}). Even though the exact value is uncertain, based on comparisons to light curves of other 03fg-like SNe~Ia (see \autoref{fig:lightcurve}), it is unlikely that the SN increased in brightness considerably after our last early-phase photometry. By comparing with SN~2012dn, the data strongly suggest a conservative limit of $M_{B,{\rm peak}} > -19.3$ mag.

\subsection{Ground-based Spectroscopy} \label{ss:spec_data}

\begin{figure*}
\begin{center}
    \includegraphics[width=3.2in]{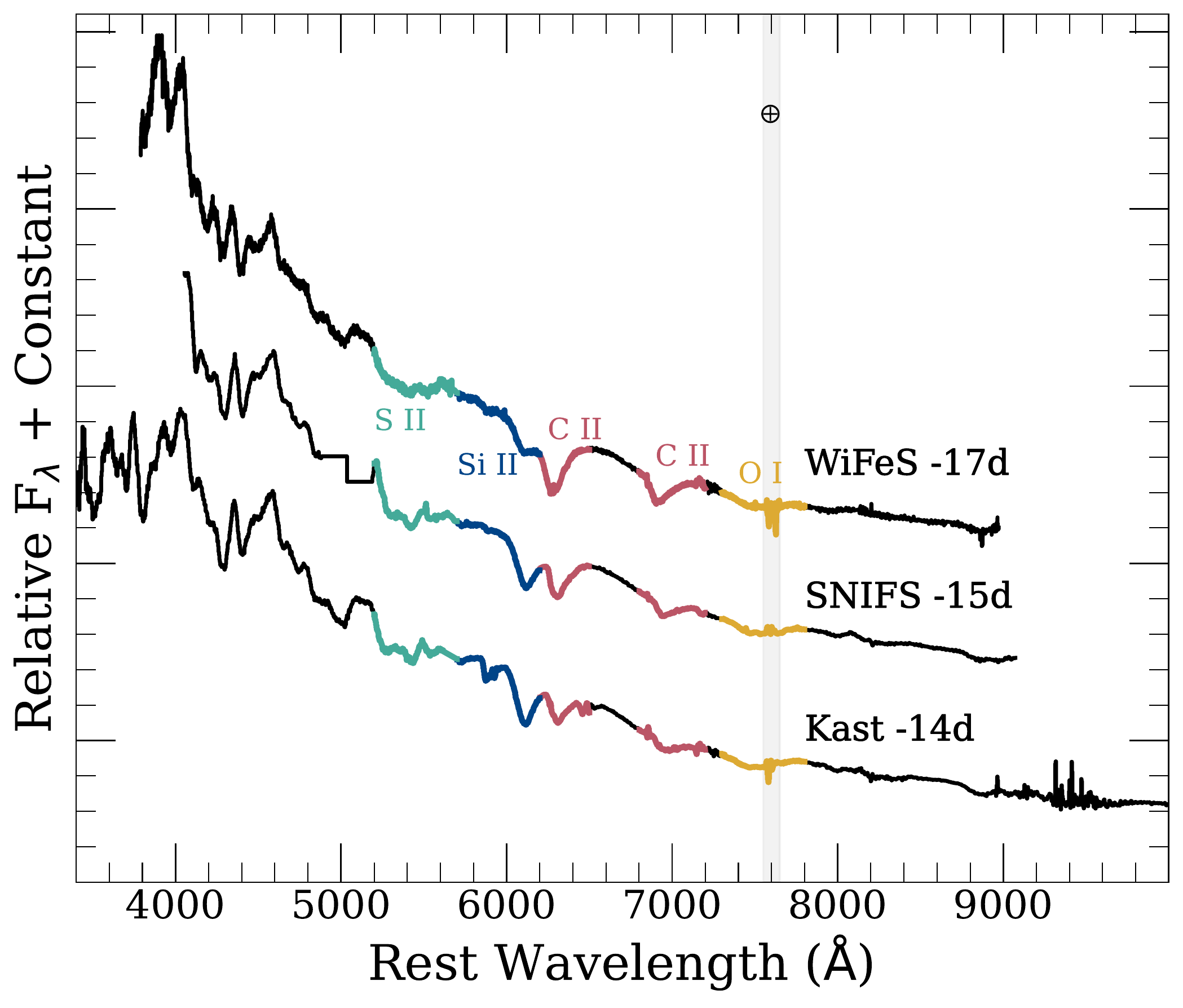}
    \includegraphics[width=3.2in]{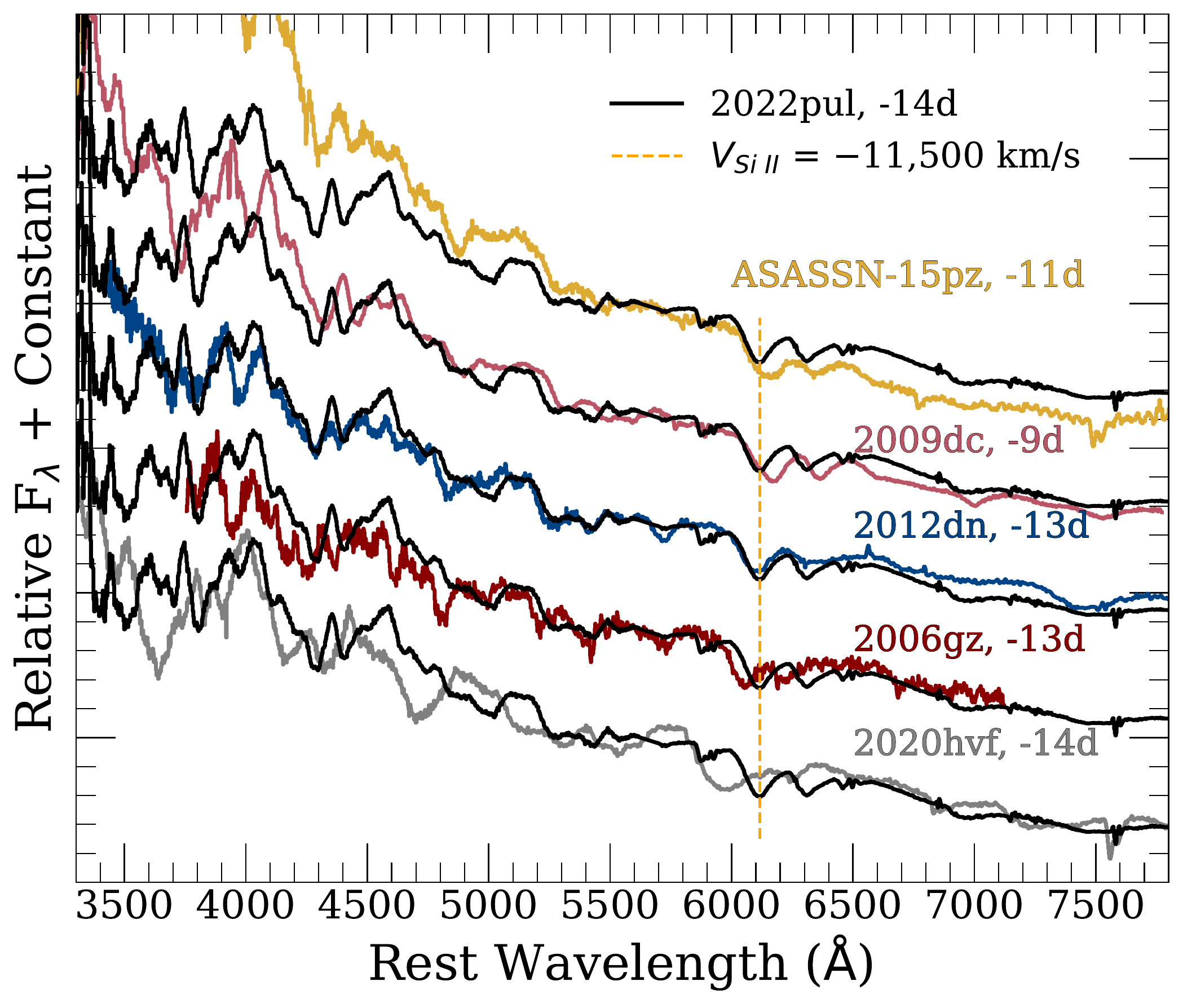}
\caption{\textit{(left)}: Early-phase optical spectra of \pul. \textit{(right)}: our $-14$ day optical spectrum of \pul\ in comparison with other 03fg-like SNe~Ia. At this phase, \pul\ is most similar to SN~2012dn. For \pul, we measure \ion{Si}{2} and \ion{C}{2} velocities of $-11{,}500$\kms and $-12{,}500$\kms, respectively.}\label{fig:spec_early}
\end{center}
\end{figure*}

\begin{figure*}
\begin{center}
    \includegraphics[width=6.8in]{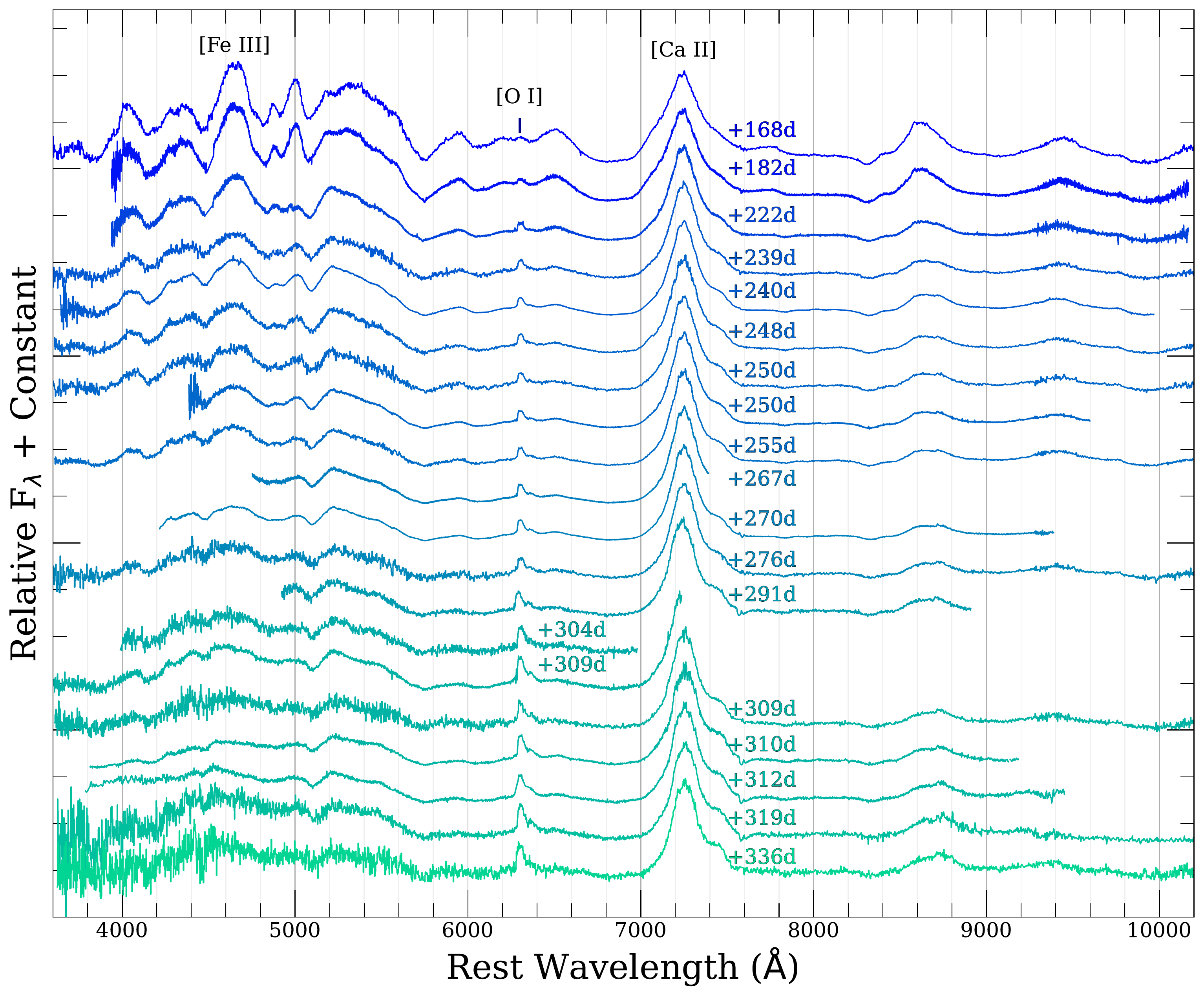}
\caption{Nebular-phase optical spectra of \pul\ spanning +168 to +336 days after peak brightness. \pul\ shows significant evolution during this period of time. In particular, [\ion{Fe}{3}] $\lambda$4701 is clearly detected 
 before $\sim 270$~d, but vanishes by the time of our last epoch. Additionally, we detect strong [\ion{Ca}{2}] $\lambda\lambda 7291$, 7324 emission and strong [\ion{O}{1}] $\lambda\lambda$6300, 6364 emission (post 240 days).}
 \label{fig:spec_time}
\end{center}
\end{figure*}

Our spectroscopic observations of \pul\ (listed in \autoref{app} in \autoref{tab:spec}) cover a range from $-17$ to $+336$ days relative to peak $B$-band brightness. The early-phase ($\leq -14$ days) optical spectra are presented in the left panel of \autoref{fig:spec_early}, and the nebular-phase ($\geq 168$ days) optical spectra are displayed in \autoref{fig:spec_time}. Ground-based NIR observations are shown in \autoref{fig:spec_comp_nir}. A complete log of spectroscopic observations is presented in \autoref{tab:spec}. All spectra have been corrected for MW extinction. 

We used the following ground-based telescopes and instruments to collect our spectroscopic sequence: the Wide-Field Spectrograph (WiFeS; \citealt{Dopita07}) on the 
Australian National University (ANU) 2.3~m telescope at the Siding Spring Observatory, the Kast spectrograph \citep{Miller93} on the Shane 3~m telescope at Lick Observatory, the Echellette Spectrograph and Imager (ESI; \citealt{Sheinis02}) mounted on the 10~m Keck~II telescope at the W.~M.~Keck Observatory, the Near-Infrared Echellette Spectrometer (NIRES; \citealt{Wilson04}) also on the Keck II telescope, the DEep Imaging Multi-Object Spectrograph (DEIMOS; \citealt{faber03}) also on the Keck II telescope, the Faint Object Camera And Spectrograph (FOCAS; \citealt{Kashikawa02} on the Subaru telescope, the Inamori-Magellan Areal Camera and Spectrograph (IMACS; \citealt{Dressler11}) on the Magellan Baade Telescope, the Goodman High Throughput Spectrograph \citep{Clemens04} on the 4.1~m Southern Astrophysical Research (SOAR) telescope, the Robert Stobie Spectrograph (RSS; \citealt{Smith06}) on the Southern African Large Telescope (SALT), the Binospec imaging spectrograph \citep{Fabricant19} on the MMT, and the Optical System for Imaging and low-Intermediate-Resolution Integrated Spectroscopy (OSIRIS) instrument mounted to the 10.4~m Gran Telescopio Canarias (GTC) at the Observatorio del Roque de Los Muchachos in La Palma.

Data from the Kast, ESI, and Goodman spectrographs were reduced using either the UCSC Spectral Pipeline\footnote{\url{https://github.com/msiebert1/UCSC_spectral_pipeline}} or the KastShiv\footnote{\url{https://github.com/ishivvers/TheKastShiv}} pipeline, which make use of standard \textsc{iraf}\footnote{IRAF is distributed by the National Optical Astronomy Observatory, which is operated by the Association of Universities for Research in Astronomy (AURA), Inc., under a cooperative agreement with the National Science Foundation (NSF).}, Python, and IDL routines for bias/overscan corrections, flat fielding, flux calibration, and telluric-line removal, using spectrophotometric standard star spectra, obtained on the same night \citep{Silverman12:bsnip}.

The DEIMOS spectra were reduced using the \textsc{DPipe} spectral reduction pipeline (created as a modified version of \textsc{LPipe}; \citealt{Perley19}) and standard \textsc{iraf} routines. Low-order polynomial fits to comparison-lamp spectra were used to calibrate the wavelength scale, and small adjustments derived from night-sky lines in the target frames were applied. The spectra were flux calibrated and telluric corrected using observations of appropriate spectrophotometric standard stars observed on the same night, at similar airmasses, and with an identical instrument configuration.

NIRES spectra were obtained as part of the Keck Infrared Transient Survey (KITS), a NASA Keck Key Mission Strategy Mission Support program (PI R. Foley). We observed the SN at two positions along the slit (AB pairs) to perform background subtraction. An A0~V star was observed immediately before or after the science observation in order to remove telluric features. We reduced the NIRES data using spextool v.5.0.2 \citep{Cushing04}; the pipeline performs flat-field corrections using observations of a standard lamp and wavelength calibration based on night-sky lines in the science data. We performed telluric correction using xtellcor \citep{Vacca03}.

IMACS data were reduced with standard IRAF routines. Flux calibration was achieved through spectra of spectrophotometric standard stars obtained during the same observing nights. Telluric correction was performed using specific standards from \citet{Bessell99}. Some residuals of saturated telluric bands are present.

Our SALT/RSS spectrum was reduced using a custom pipeline based on standard Pyraf \citep{Pyraf} spectral reduction routines and the PySALT package \citep{Crawford10}; we removed cosmic rays, host-galaxy lines and continuum, and telluric absorption. MMT/Binospec data were reduced using the Binospec IDL pipeline \citep{Kansky19}.

The Subaru/FOCAS spectrum was reduced following standard procedures with IRAF, including bias subtraction, flat-fielding, cosmic-ray rejection, sky subtraction, one-dimensional (1D) spectral extraction, wavelength calibration using ThAr lamp and night-sky lines, and flux calibration with the standard star Feige~34 \citep[e.g.,][]{maeda22}. 

OSIRIS observations were conducted using the R1000B and R1000R grisms with wavelength ranges 3630--7500~\AA\ and 5100--10,000~\AA, respectively. The reduction process was performed using version 1.11.0 of \textit{PypeIt} \citep{Prochaska20}\footnote{\url{https://github.com/pypeit/PypeIt}}

For absolute flux calibration, we matched the ground based optical spectra to available photometry. We lacked near-infrared photometry of \pul, so the Keck/NIRES data use spectroscopic flux standards only for calibration.

\subsection{\textit{JWST} Spectroscopy} 

\begin{deluxetable}{lll}
    \tablecaption{JWST SN~2022pul Observation Details\label{tab:details}}
    \tablehead{\colhead{Setting} & \colhead{NIR} & \colhead{MIR}}
    \tablecolumns{3}
        \startdata
        Instrument & NIRSpec & MIRI \\
        Mode & FS & LRS \\
        Wavelength Range & 0.6 $-$ 5.3$\mu$m\ & 5--14~$\mu$m\ \\
        Slit & S200A1 ($0.2'' \times 3.3''$) & Slit \\
        Grating/Filter & PRISM/CLEAR & P750L \\
        R $=\lambda/\Delta\lambda$ & $\sim$100 & $\sim 40$--250 \\
        Subarray & FULL & FULL \\
        Readout Pattern & NRSIRS2RAPID & FASTR1 \\
        Groups per Integration & 10 & 89 \\
        Integrations per Exposure & 2 & 5 \\
        Exposures/Nods & 3 & 2 \\
        Total Exposure Time & 963 s & 2492 s \\
        Target Acq. Exp. Time & 14 s & 89 s
        \enddata
\end{deluxetable}%\vspace{-24pt}

\begin{figure*}[htb!]
    \centering
    \includegraphics[width=\textwidth]{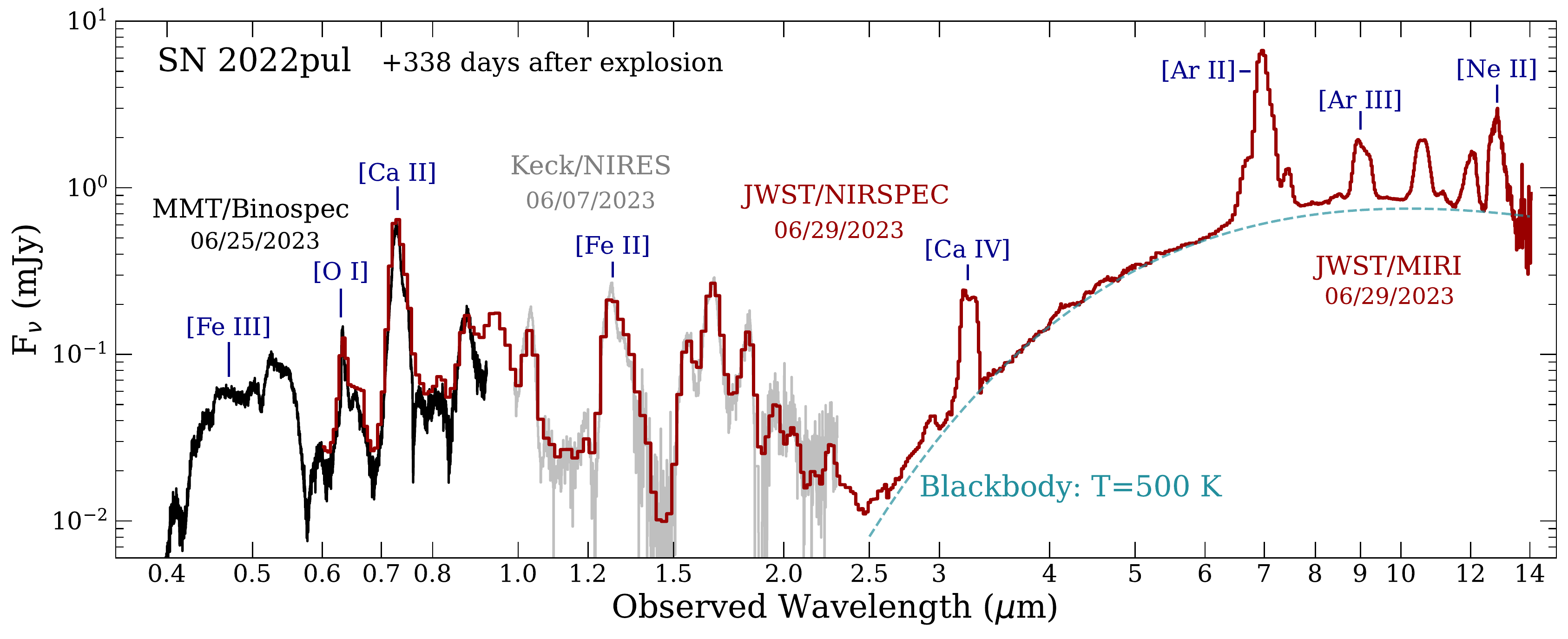}
    \caption{Combined optical $+$ NIR $+$ MIR spectrum of the 03fg-like Type Ia SN 2022pul in the nebular phase at $+338$ after explosion. With optical data from MMT/Binospec at a similar phase, along with slightly earlier NIR data from Keck/NIRES, our \textit{JWST}/NIRSpec and \textit{JWST}/MIRI data continuously cover 0.35--14~$\mu$m, and are the first MIR spectra of an 03fg-like SN Ia. The NIRSpec flux is unscaled and matches with the optical and MIR spectra which have been scaled to the photometry. The Keck/NIRES spectrum has been rebinned to lower resolution for presentation purposes.
    }\label{fig:full_spec}
\end{figure*}

%%%%%% Text by Lindsey:  %%%%%%%%%%%%%%%%%
We observed SN 2022pul in the nebular phase on 2023 June 29 at 338 rest-frame days post-explosion with \textit{JWST}, using both the Near Infrared Spectrograph (NIRSpec) in the Fixed Slits (FS) Spectroscopy mode \citep{Jakobsen2022,Birkmann2022,Rigby2022} and MIRI in the Low Resolution Spectroscopy (LRS) mode \citep{Kendrew2015,Kendrew2016,Rigby2022}. These data are presented in \autoref{fig:full_spec}. Our NIRSpec observations used the S200A1 ($0.200''$ wide $\times$ $3.300''$ long) slit with the PRISM grating and CLEAR filter, and our MIRI observations used the LRS slit with the P750L disperser. Together the \textit{JWST} spectra continuously span 0.6--14~$\mu$m. Since the fluxes of the NIRSpec and MIRI spectra at 5~$\mu$m agree with no scaling or modification, we choose to transition from the NIRSpec spectrum to the MIRI spectrum at this wavelength. \autoref{tab:details} gives further details of the \textit{JWST} observation settings.

We reduced the \textit{JWST} data using the publicly available ``jwst''\footnote{\url{https://github.com/spacetelescope/jwst}} pipeline \citep[version 1.11.1;][]{Bushouse_JWST_Calibration_Pipeline_2022} routines for bias and dark subtraction, background subtraction, flat-field correction, wavelength calibration, flux calibration, rectification, outlier detection, resampling, and spectral extraction. To check the reduction of the automatic pipeline, we re-extracted both spectra by manually running Stage 3 of the pipeline (calwebb\_spec3) from the Stage 2 (calwebb\_spec2) data products; we found excellent agreement between the extractions so we use the automated pipeline reductions available on Mikulski Archive for Space Telescopes (MAST)\footnote{\url{https://mast.stsci.edu/portal/Mashup/Clients/Mast/Portal.html}}. Upon inspection of the two-dimensional (2D) spectral image, the NIRSpec spectrum suffered from several reduction artifacts (potentially cosmic rays or bad pixels) not removed by the automated pipeline, so we manually clip these artifacts out in the spectrum. We take special care to remove a single-pixel artifact on the red side of the 3.2~$\mu$m feature to preserve the underlying line profile shape.

The MIRI/LRS slit mode wavelength calibration has a known issue causing a 0.02--0.05~$\mu$m uncertainty which is largest at the shortest wavelengths\footnote{Details at \url{https://jwst-docs.stsci.edu/jwst-calibration-pipelinecaveats/jwst-miri-lrs-pipeline-caveats}}. \citet{Beiler2023} use observations of a Y-dwarf to make a further empirical correction to the wavelength solution given by 
\begin{equation*}
    \Delta \lambda = 0.0106~\lambda_{\text{MIRI}} - 0.120\,\mu\text{m}.    
\end{equation*}
We applied this additional wavelength correction to our MIRI spectrum.

From the LRS verification image of \pul, we measured MIRI F1000W photometry of $F_\nu = 1.130 \pm 0.003$~mJy, corresponding to $16.267 \pm 0.003$ mag AB \citep{Oke83}. The photometry was done on the F1000W data from the \textit{JWST} pipeline using a 70\% encircled energy aperture radius (4.3 pixels) and inner and outer sky radii of (respectively) 6.063 and 10.19 pixels (and a corresponding aperture correction was also applied). The measured photometry agreed with the flux from integrating the MIRI/LRS spectrum of \pul\ over the F1000W passband to within 7\%, and we rescale the MIRI spectrum flux to match the photometry precisely. The NIRSpec spectrum does not have a verification image to measure photometry and check the spectral flux calibration; however, it matches both the optical and MIR spectra well, so we do not adjust its flux calibration.

\section{Analysis}\label{s:anal}

\subsection{Spectroscopic Evolution}\label{sec:time}

\begin{figure}
\begin{center}
    \includegraphics[width=3.2in]{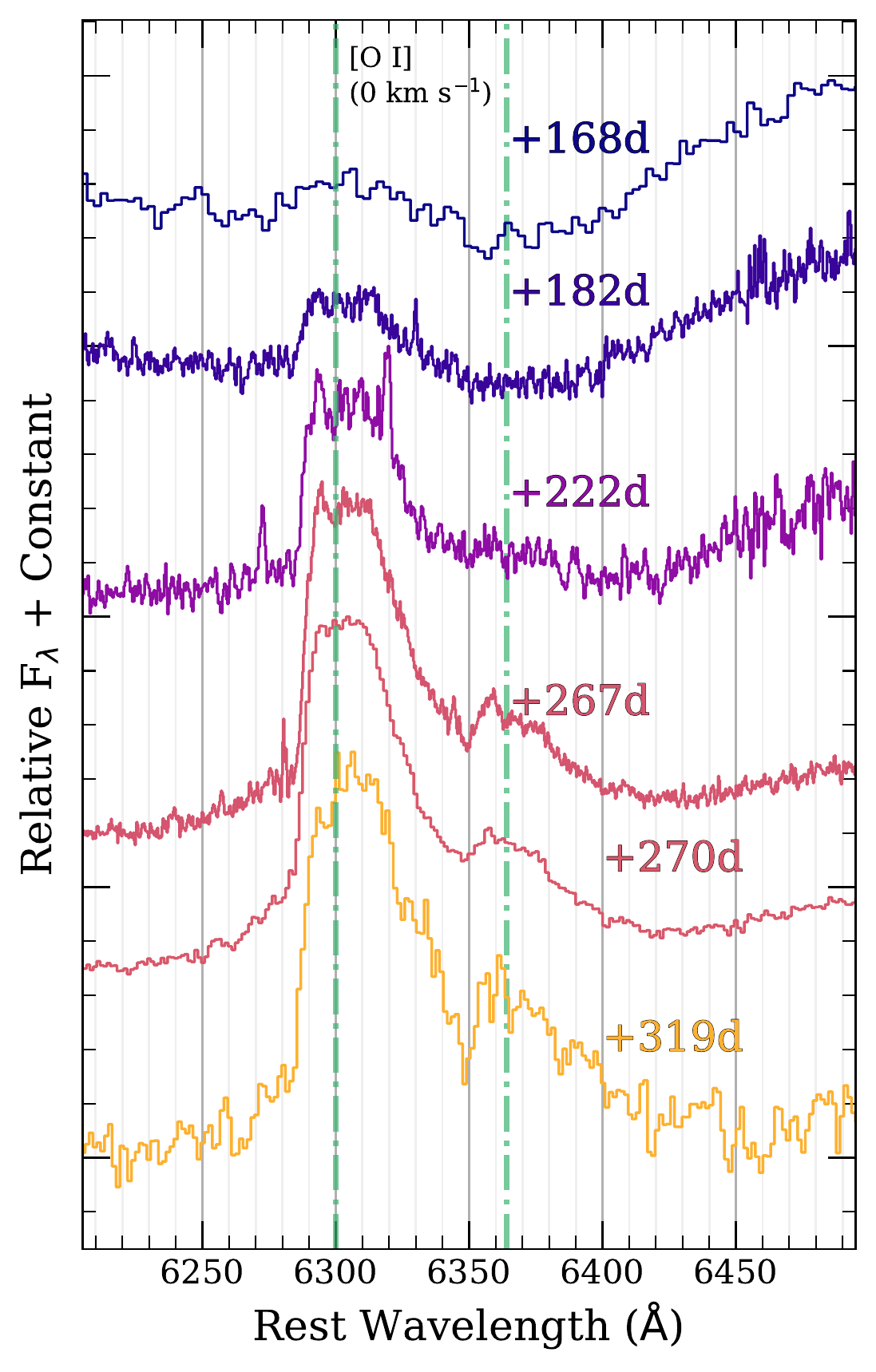}
\caption{Selected spectra from \autoref{fig:spec_time} zoomed-in near [\ion{O}{1}]. After 222 days, strong components of both [\ion{O}{1}] $\lambda$6300 and $\lambda$6364 are detected. [\ion{O}{1}] $\lambda$6300 is asymmetric with a steeper blueshifted edge and has FWHM $\approx 2000$~km~s$^{-1}$. Both emission profiles are flat-topped and centered at 300~km~s$^{-1}$ (0~km~s$^{-1}$ green-dashed line for reference). }\label{fig:spec_time_OI}
\end{center}
\end{figure}

\begin{figure*}
\begin{center}
    \includegraphics[height=3.8in]{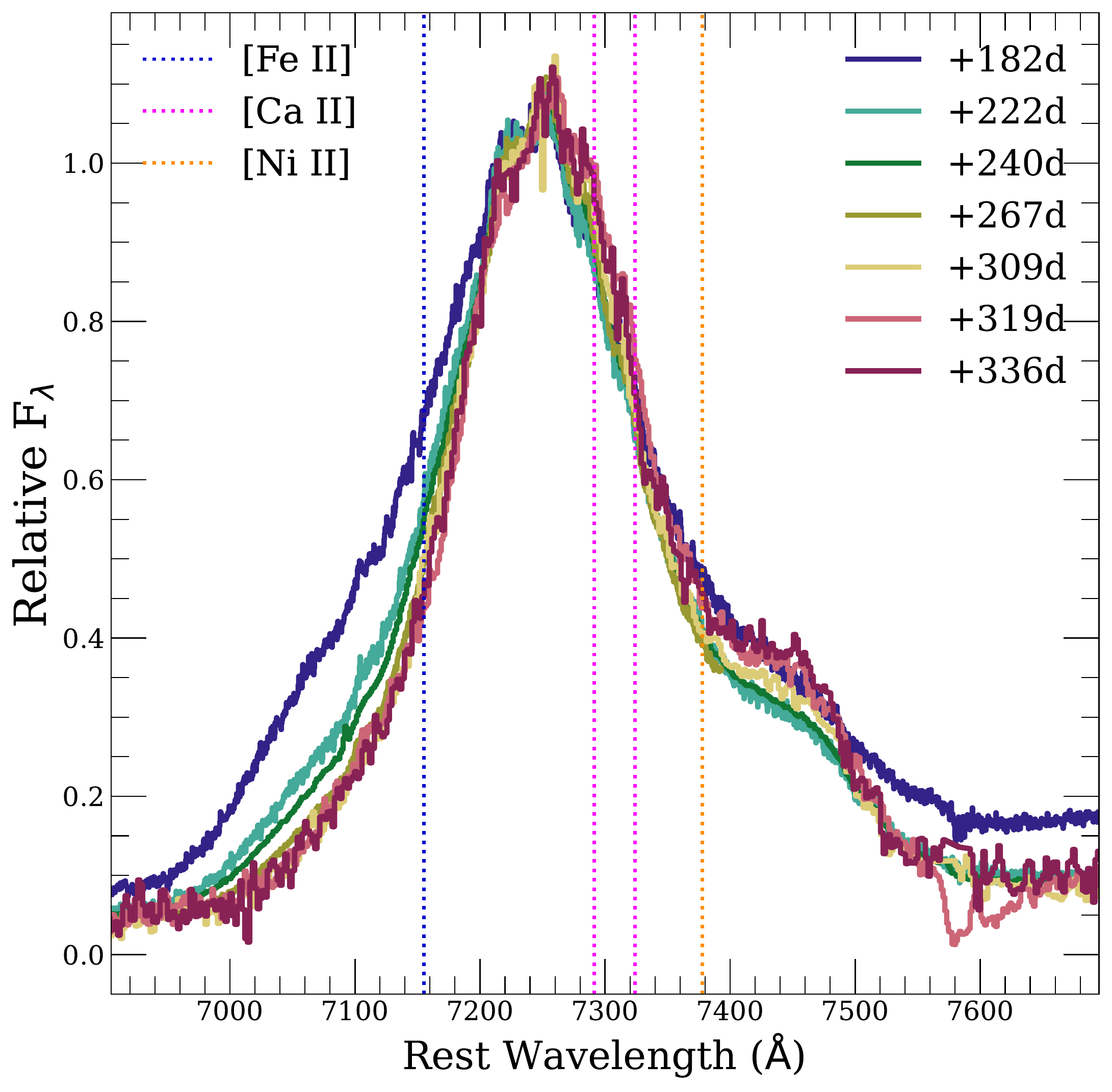}
    \includegraphics[height=3.8in]{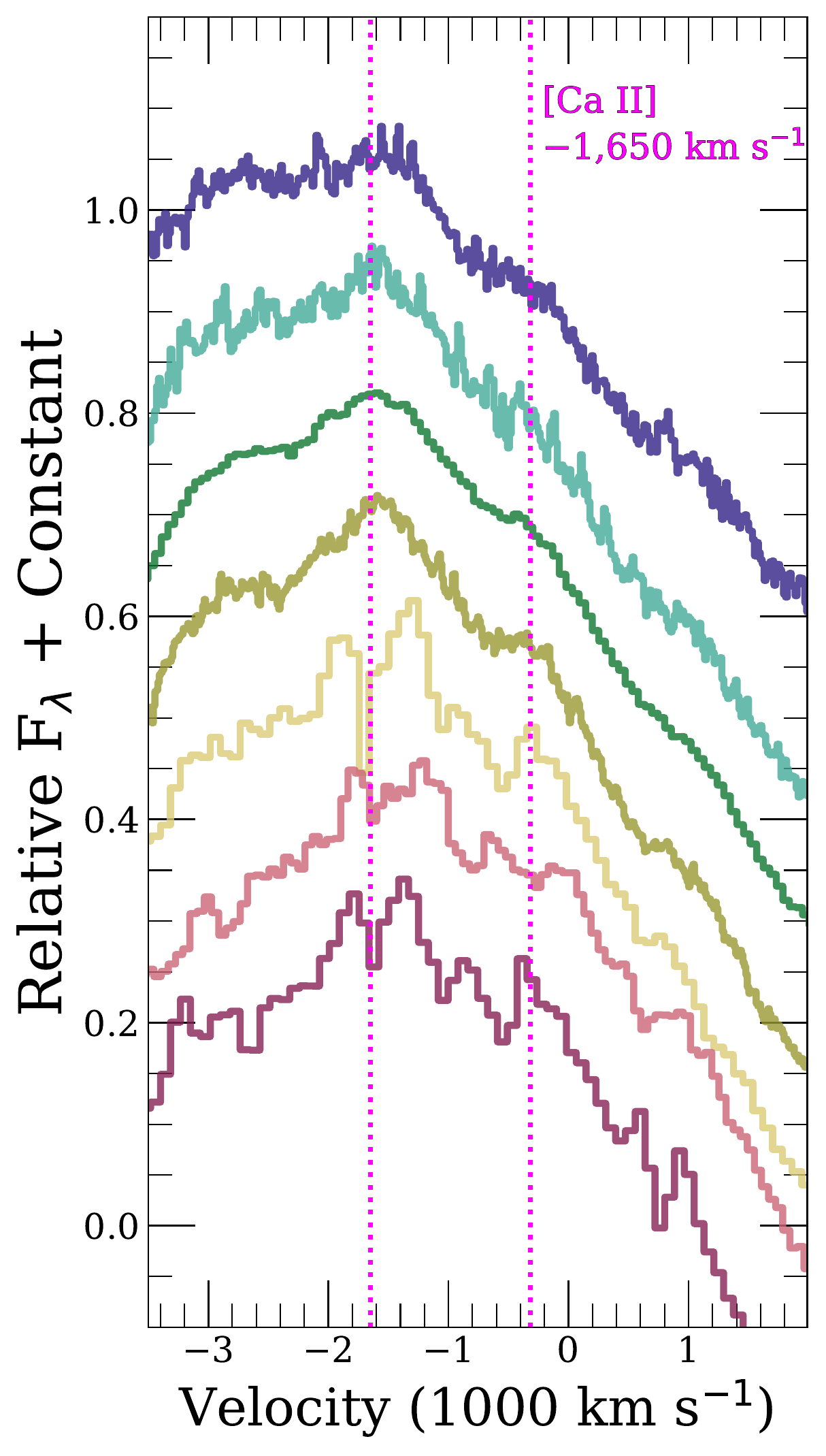}
\caption{\textit{(left):} Selected spectra from \autoref{fig:spec_time} zoomed-in on the $7{,}300$~\AA\ emission complex and normalized to its peak. The strongest emission lines of [\ion{Fe}{2}], [\ion{Ca}{2}], and [\ion{Ni}{2}] are shown at their rest wavelengths for reference (blue, pink, and orange dashed vertical lines, respectively). \textit{(right):} The same spectra zoomed-in near the peak of [\ion{Ca}{2}] and offset by constant values. The [\ion{Ca}{2}] emission is blueshifted by $\sim$1{,}700~km~s$^{-1}$ (pink dashed vertical lines for reference). }\label{fig:spec_time_CaII}
\end{center}
\end{figure*}

Our early photospheric spectra of \pul\ show several of the characteristic features of 03fg-like SNe~Ia. In the left panel of \autoref{fig:spec_early} we present our early-phase optical spectra of \pul, and in the right panel we compare our $-14$ day optical spectrum of \pul\ to other 03fg-like SNe. Consistent with our comparisons to nebular spectra in Section \ref{s:spec_comp}, we find that \pul\ is most similar to SN~2012dn in the photospheric phase. We measure an \ion{Si}{2} absorption velocity of $-11{,}500$~\kms\ ($-11{,}700$~\kms\ for SN~2012dn). We also measure strong \ion{C}{2} absorption at this phase, which is likely unburned carbon in the outer layers of the ejecta (consistent with other 03fg-like SNe). The higher velocity and low luminosity relative to the observed population 03fg-like SNe \citep{Ashall21} could indicate that \pul\ is the result of a lower total mass explosion.

\pul\ is remarkably well observed at optical wavelengths during its transition into the nebular phase. In \autoref{fig:spec_time}, we present data from a variety of ground-based sources (described in \autoref{ss:spec_data}) ranging from +168 to 336 days after peak $B$-band brightness, highlighting the evolution that occurs in the nebular emission lines during this phase.

In our earliest nebular spectrum (+168 days), emission from [\ion{Fe}{3}] $\lambda$4701 is clearly present. This feature proceeds to weaken with time and is absent from the spectrum by +309 days. At $>182$ days we detect emission from [\ion{O}{1}] $\lambda\lambda$6300, 6364, and in all spectra we detect strong emission from [\ion{Ca}{2}] $\lambda \lambda7291$, 7324. The strength of [\ion{O}{1}] increases relative to [\ion{Ca}{2}] over this time period with the [\ion{Ca}{2}]/[\ion{O}{1}] peak-intensity ratio changing from 3.5 to 2.7. This ratio is similar to those observed in Ca-rich SNe and SNe~II at $>100$ days after peak brightness \citep{Valenti14,Milisavljevic17,Prentice22}. Additionally, the strength of the emission component of the \ion{Ca}{2} NIR triplet weakens with time relative to [\ion{Ca}{2}]. The \ion{Ca}{2} NIR triplet has a photospheric component with blueshifted absorption near $\sim -$12{,}000~km~s$^{-1}$ at all phases. 

In \autoref{fig:spec_time_OI}, using a selection of spectra from \autoref{fig:spec_time}, we display the evolution of the [\ion{O}{1}] emission. At all phases after +222 days, there is clear emission from both [\ion{O}{1}] $\lambda$6300 and $\lambda$6364. The ratio of [\ion{O}{1}] $\lambda$6300 to [\ion{O}{1}] $\lambda$6364  in our +270 day IMACS spectrum is $3.0 \pm 0.1$, consistent with what we would expect given the measured oscillator strengths of these lines ([\ion{O}{1}] $\lambda$6300/ [\ion{O}{1}] $\lambda$6364 = 3.1). [\ion{O}{1}] might be present at +168 days, but since the continuum is significantly different at that phase, it is hard to say definitively. From +182 to +319 days, the peak of the [\ion{O}{1}] emission is slightly redshifted ($\sim 300$~km~s$^{-1}$) and appears to evolve from a flat-topped profile to a more Gaussian profile. Furthermore, this emission is asymmetric with a steeper blue edge and more shallow-sloped red edge. This kind of ``sawtoothed" emission has been observed in the nebular-line profiles of other 03fg-like SNe \citep{Taubenberger13:sc, Siebert23a}. From our highest S/N spectrum we measure a [\ion{O}{1}] $\lambda$6300 full width at half-maximum intensity (FWHM) $\approx 2{,}000$~km~s$^{-1}$ (Paper II: \citealt{Kwok23b}). 

In addition to its strength, the morphology of the 7300~\AA\ emission complex evolves during the nebular phase. In the left panel of \autoref{fig:spec_time_CaII}, we display nebular spectra normalized to the peak of this feature ranging from +182 to 336 days. Vertical dashed lines represent the rest wavelength of the strongest lines of [\ion{Fe}{2}], [\ion{Ca}{2}], and [\ion{Ni}{2}]. Over 137 days, we observe a gradual increase in the slope of the blue edge of this feature and slight redward shift of the peak. The red ``shoulder" also becomes more pronounced over time. 

In the right panel of \autoref{fig:spec_time_CaII}, we show the same spectra zoomed-in on the peak emission from [\ion{Ca}{2}] and offset by constant values. Several 03fg-like SNe show evidence for additional narrower velocity components of [\ion{Ca}{2}] \citep{Siebert23a}. In a few of our spectra, there is additional structure centered near $-1{,}650$~km~s$^{-1}$ (pink dashed line for reference). This structure could be coming from another component of [\ion{Ca}{2}], but the resolution of our spectra limits our ability to characterize its shape. 

The evolution of several of the emission features in the nebular phase of \pul\ shows the necessity for high-quality spectral time series of these events. In the following section, we examine the similarities of \pul\ with other SNe. 

\subsection{Spectroscopic Comparisons}\label{s:spec_comp}

\begin{figure*}
\begin{center}
    \includegraphics[width=6.8in]{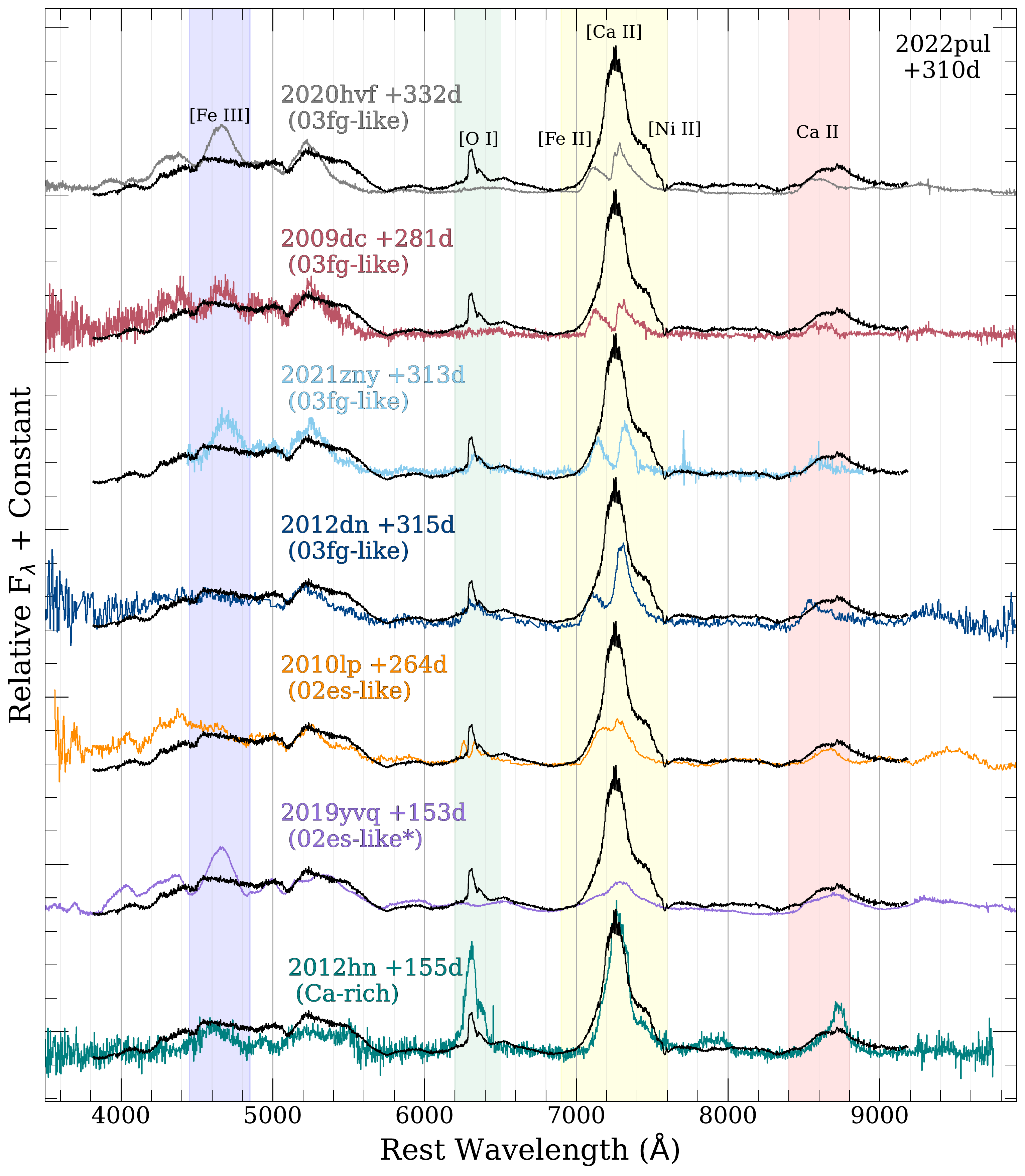}
\caption{Optical spectrum of SN~2022pul (black curves) at +310~days after peak brightness compared with those of other peculiar SNe~Ia in the nebular phase. From top to bottom, we show SN~2020hvf (gray), SN~2009dc (red), SN~2021zny (light blue), and SN~2012dn (blue), which are all 03fg-like SNe~Ia; SN~2010lp (orange), a peculiar SN~2002es-like SN which had nebular [\ion{O}{1}] emission (orange); SN~2019yvq (purple), a peculiar SN~Ia with strong [\ion{Ca}{2}] (classified as ``transitional 02es-like" by \citealt{Burke21});  and SN~2012hn (teal), a Ca-rich SN with broad [\ion{O}{1}] and [\ion{Ca}{2}] features. Where relevant, we have clipped narrow emission lines from the host galaxy for better visualization. Several spectral regions are highlighted: [\ion{Fe}{3}] $\lambda4701$ (blue); [\ion{O}{1}] $\lambda\lambda$6300, 6364 (green); the feature at 7300~\AA\ which includes possible contributions from [\ion{Fe}{2}] $\lambda7155$, [\ion{Ni}{2}] $\lambda7378$, and [\ion{Ca}{2}] $\lambda\lambda 7291$, 7324 (yellow); and the \ion{Ca}{2} NIR triplet (red).}\label{fig:spec_comp}
\end{center}
\end{figure*}

\begin{figure}[htb!]
\begin{center}
    \includegraphics[width=2.55in]{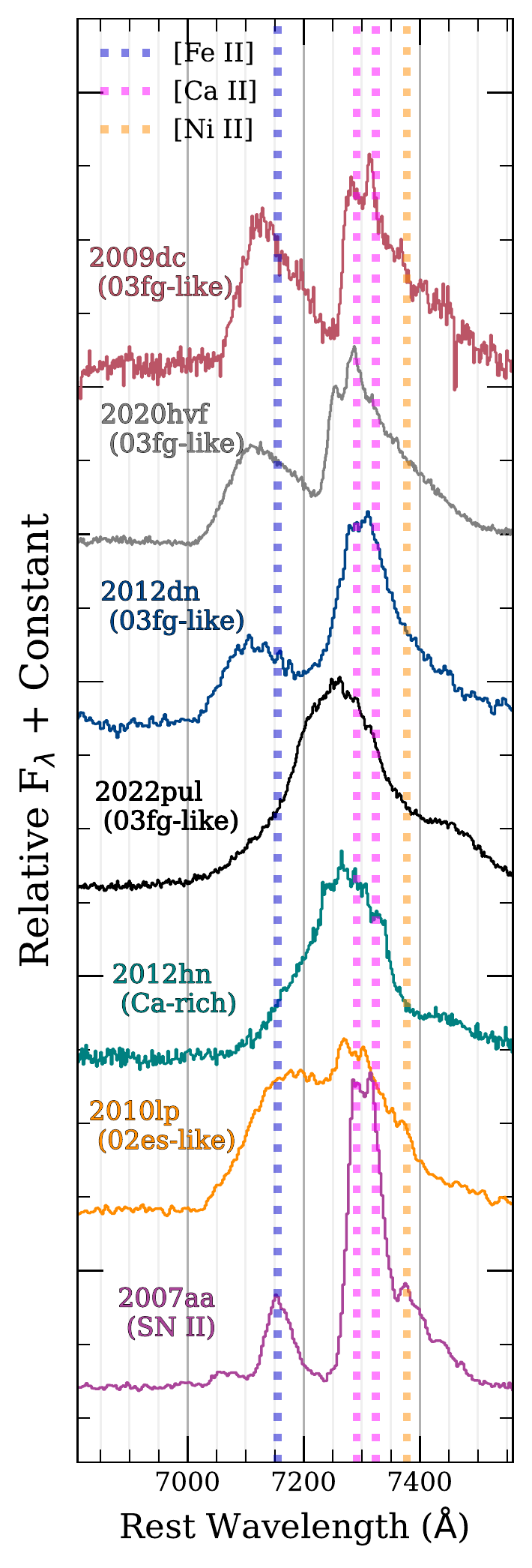}
\caption{The 7300~\AA\ emission feature of \pul\ in comparison with the nebular spectra of other peculiar SNe~Ia from \autoref{fig:spec_comp}. From top to bottom the SNe displayed are SN~2009dc (03fg-like, red curve), SN~2020hvf (03fg-like, gray curve), SN~2012dn (03fg-like, blue curve), \pul\ (03fg-like, black curve), SN~2012hn (Ca-rich SN, green curve), SN~2010lp (``02es-like," orange curve), and SN~2007aa (SN II, purple curve).}\label{fig:spec_comp_7300}
\end{center}
\end{figure}

\begin{figure}
\begin{center}
    \includegraphics[width=3.2in]{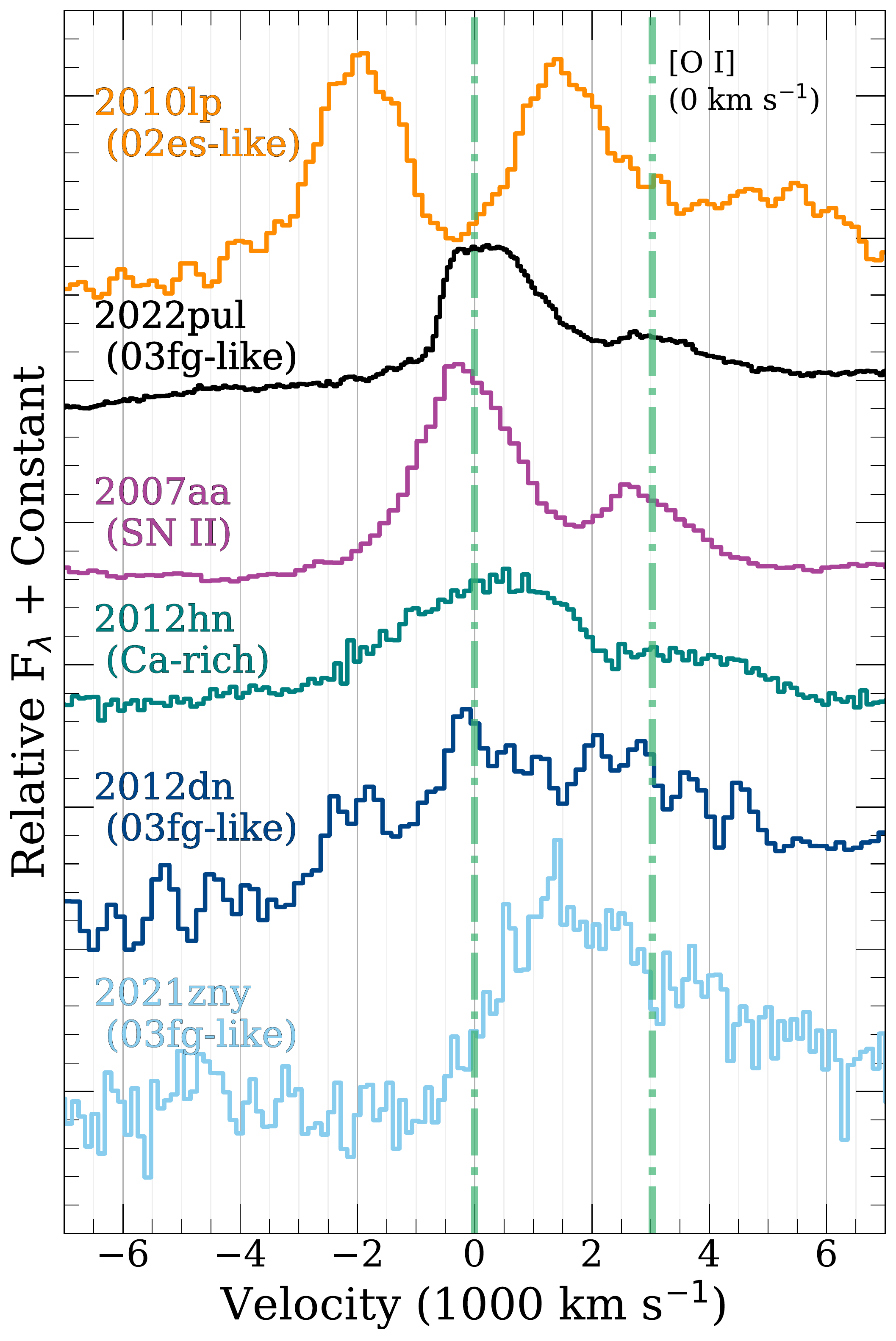}
\caption{Comparison of the nebular Subaru spectrum of \pul\ (+240~d, black) with other SNe~Ia that have [\ion{O}{1}] emission. Starting from the top, the other SNe are the ``02es-like" SN~2010lp, the SN~II 2007aa, the Ca-rich SN~2012hn, the 03fg-like SN~Ia 2012dn, and the 03fg-like SN~Ia 2021zny \citep{Dimitriadis23}. The dashed green lines show [\ion{O}{1}] $\lambda\lambda$6300, 6364 at $0$~km~s$^{-1}$. The velocity and width of [\ion{O}{1}] in \pul\ is most similar to those seen in SNe~II and Ca-rich SNe.}\label{fig:spec_comp_OI}
\end{center}
\end{figure}

The nebular features of \pul\ add to the growing diversity present in the 03fg-like subclass of SNe~Ia, and exhibit interesting similarities to other types of peculiar SNe~Ia. In \autoref{fig:spec_comp}, we show our +310 day nebular spectrum of \pul\ in comparison with four other 03fg-like SNe~Ia (SN~2020hvf, \citealt{Siebert23a}; SN~2009dc, \citealt{Silverman11:09dc}; SN~2021zny, \citealt{Dimitriadis23}; and SN~2012dn, \citealt{Taubenberger19}), two 02es-like SNe~Ia (SN~2010lp, \citealt{Kromer13}; and SN~2019yvq, \citealt{Siebert20b}), and one Ca-rich SN (SN~2012hn, \citealt{Valenti14,Lyman14:carich}).

The 03fg-like SNe shown here demonstrate the varying ionization levels observed in this subclass. Of these SNe, both \pul\ and SN~2012dn lack emission from [\ion{Fe}{3}]. Stronger [\ion{Fe}{3}] emission in SN~2009dc, SN~2020hvf, and SN~2021zny at similar phases show that the ionization states of 03fg-like SNe can evolve on varying timescales, unlike normal SNe~Ia. This could be related to the mass of $^{56}$Ni generated in the explosion, or the varying recombination rates caused by different densities. SN~2009dc and SN~2020hvf had higher peak brightnesses ($M_B = -20.3$~mag and $M_B = -19.9$~mag, respectively), while SN~2012dn and \pul\ are fainter relative to the population of other 03fg-like SNe ($M_B = -19.28$~mag and $M_B = -18.9$~mag, respectively). 

The most striking nebular features of \pul\ are the strong [\ion{O}{1}] and [\ion{Ca}{2}] emission. So far, only two previous 03fg-like SNe display convincing [\ion{O}{1}] emission, SN~2012dn and SN 2021zny \citep{Taubenberger19, Dimitriadis23}; however, in both cases the emission is significantly broader (see \autoref{fig:spec_comp_OI}). The width of [\ion{O}{1}] in \pul\ is more similar to that seen in the much lower luminosity ($M_B = -17.7$~mag) 02es-like SN~2010lp. The [\ion{Ca}{2}] emission is also the broadest observed in an 03fg-like SN~Ia. Relative to [\ion{Fe}{3}], the strength and width of this feature are most similar to the Ca-rich SN~2012hn. 

\subsection{Ca II Emission}\label{sec:CaII}

The [\ion{Ca}{2}] emission is explored in detail in Figures \ref{fig:spec_comp_7300}. Here we compare the 7300~\AA\ emission profile of \pul\ (black curve) with other 03fg-like SNe (SN~2009dc, SN~2020hvf, and SN~2012dn), a Ca-rich SN~2012hn, the low-luminosity 02es-like SN~2010lp, and SN~II~2007aa. Rest wavelengths of [\ion{Fe}{2}], [\ion{Ca}{2}], and [\ion{Ni}{2}] are shown with blue, pink, and orange dashed vertical lines, respectively. 

The broad [\ion{Ca}{2}] present in \pul\ is significantly affected by line overlap, complicating the interpretation of this feature. The other 03fg-like SNe presented here all show ``sawtooth"-shaped emission in either one or both of the profiles of [\ion{Fe}{2}] and [\ion{Ca}{2}]. Given that an asymmetric [\ion{Fe}{2}] line profile is present in the NIR and MIR (see \autoref{fig:spec_comp_nir}, and \citealt{Kwok23b}), we should expect a similar underlying component in this feature of \pul. Furthermore, the emission peaks of [\ion{Ca}{2}] observed in 03fg-like SNe all appear blueshifted. If the dust is formed in the SN \citep{Johansson23} and is obscuring part or all of the ejecta, one would expect the redshifted emission to be more heavily extinguished. This could give rise to the trend of the preferentially blueshifted [\ion{Ca}{2}] emission observed in these SNe. 03fg-like SNe (and SN~2010lp) also display narrow components of [\ion{Ca}{2}] and a similar component could be present in \pul\ (see the right panel of \autoref{fig:spec_time_CaII}).  

The shoulder on the red side of the 7300~\AA\ feature could be emission from [\ion{Ni}{2}], but the emission profiles of the strong intermediate-mass elements (IMEs) seen in the MIR \citep{Kwok23b} might indicate that this is instead a contribution from a broader flat-topped component of [\ion{Ca}{2}]. 

The broad [\ion{Ca}{2}] is more similar to what is seen in Ca-rich SNe, which also have low-ionization nebular-phase spectra and [\ion{O}{1}] emission. We estimate the broad component of [\ion{Ca}{2}] in \pul\ to have FWHM $=7{,}300$~\kms, compared to $5{,}500$~\kms for that of the Ca-rich SN~2012hn \citep{Prentice22}, $5{,}600$~\kms for the peculiar 02es-like SN~2019yvq \citep{Siebert20b}, and $1{,}450$~\kms that we measure for SN~2012dn.

Lastly, since both [\ion{O}{1}] and [\ion{Ca}{2}] emission are typically present in the nebular phase of core-collapse SNe, we compare the 7300~\AA\ emission of \pul\ to that of SN~II 2007aa. The widths of these features vary greatly among SNe~II, but if we select an SN with similar [\ion{O}{1}] emission (see \autoref{fig:spec_comp}), we see that its [\ion{Ca}{2}] emission is quite narrow in comparison with that of \pul.

\subsection{O I Emission}\label{sec:OI}

In \autoref{fig:spec_comp_OI}, we compare the [\ion{O}{1}] emission profile in our highest signal-to-noise ratio (S/N) spectrum to those of other SNe where this emission feature is also detected. As previously discussed in Section \ref{sec:time}, the [\ion{O}{1}] emission in \pul\ is offset by 300~\kms, flat-topped, and has two clear profiles corresponding to [\ion{O}{1}] $\lambda$6300 and  [\ion{O}{1}] $\lambda$6364. We observe the most similar emission in an SN~II, SN~2007aa at +336 days (purple curve), having a similar velocity and line ratio. SN~2010lp also shows clear and narrow [\ion{O}{1}] emission, but offset by $\sim -1{,}800$~\kms. \citet{Taubenberger13:10lp} attribute this type of emission as evidence for a violent merger like that presented by \citet{Pakmor12} where unburned oxygen (or oxygen from carbon burning) from the secondary WD is present at low velocities but is not macroscopically mixed in velocity space as would be expected for a Chandrasekhar-mass deflagration \citep{Roepke07}. The spectral models of \citet{Mazzali22} also find that SN~2010lp was consistent with a violent merger. The similar emission seen in \pul\ could point toward a similar progenitor scenario; however, SN~2010lp was significantly lower-luminosity than \pul. More modeling is needed to understand if violent mergers can produce the luminosities observed and nickel masses inferred for 03fg-like SNe. 

As shown in SN~2012dn and SN~2021zny (\autoref{fig:spec_comp_OI}), [\ion{O}{1}] emission has been observed in other 03fg-like SNe (see \citealt{Dimitriadis23}). Broad [\ion{O}{1}] has also been observed in Ca-rich SNe like 2012hn (green curve). \citealt{Zenati+23_carich} found that the merger of a hybrid He-C/O WD with a lower mass C/O WD could explain some of the observed properties of Ca-rich transients. Interestingly, these SNe also tend to be found in remote locations of their host galaxies \citep{Kasliwal12, Foley15:carich}. 

\subsection{NIR Emission}\label{sec:nir}

\begin{figure}
\begin{center}
    \includegraphics[width=3.2in]{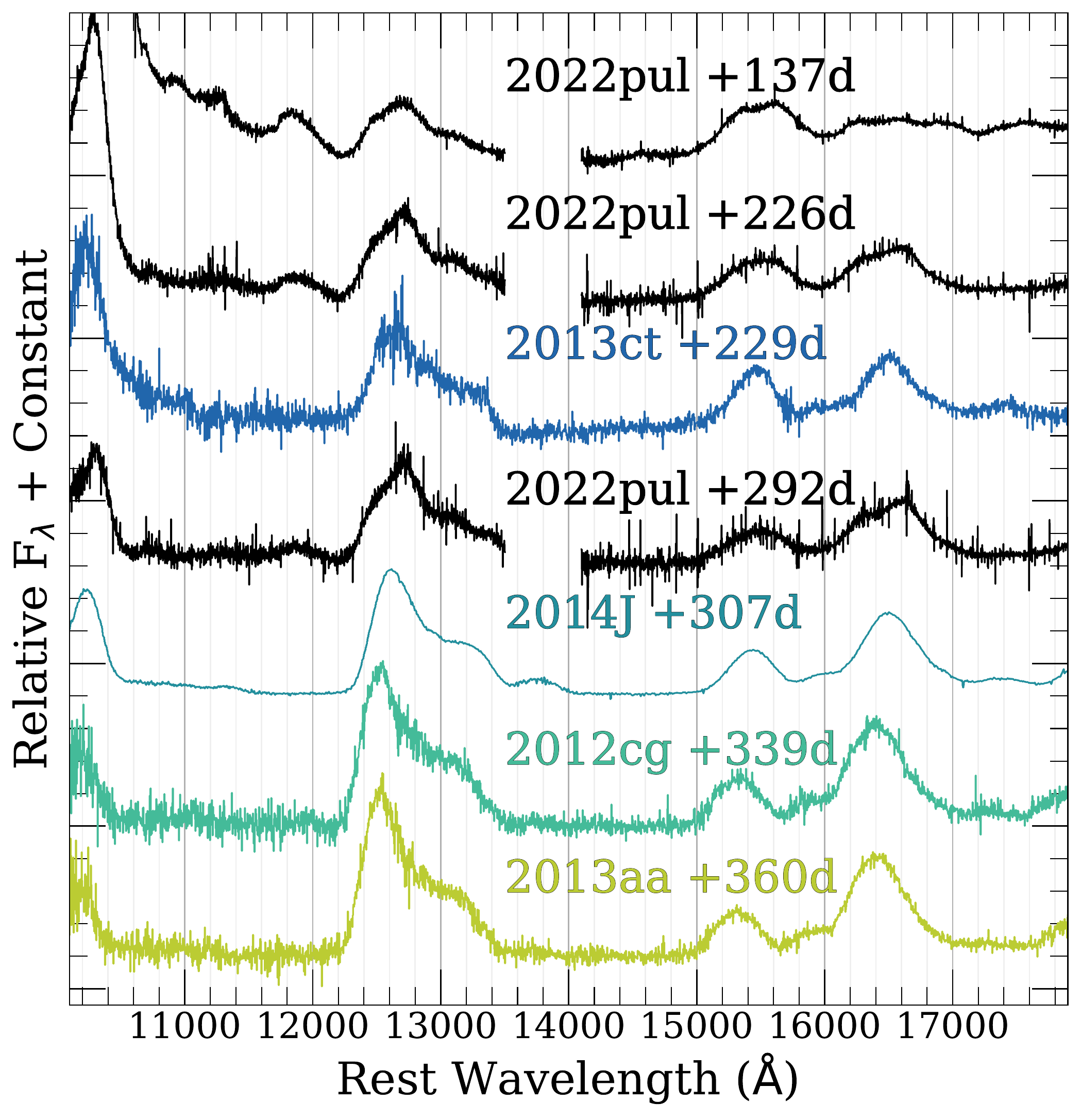}
\caption{Comparison of our nebular-phase NIR spectra of \pul\ (black curves) to those of other SNe~Ia at similar phases. The [\ion{Fe}{2}] (at $12{,}600$~\AA\ and $16{,}400$~\AA) and [\ion{Co}{2}] ($15{,}500$~\AA) profiles are more asymmetric than those seen in normal SNe~Ia. }\label{fig:spec_comp_nir}
\end{center}
\end{figure}

Since \textit{JWST} prism spectroscopy has relatively low resolution in the NIR, ground-based NIR spectroscopy obtained with NIRES provides critical constraints on the distribution of iron-group elements (IGEs) in the SN ejecta. In \autoref{fig:spec_comp_nir}, we compare our nebular NIR spectra with those of other normal SNe~Ia from \citet{Maguire18} and SN~2014J from \citep{Diamond18}. The [\ion{Fe}{2}] 1.26~$\mu$m and [\ion{Fe}{2}] 1.64~$\mu$m features in \pul\ have significantly different morphologies than those observed in other normal SNe~Ia (colored curves). As shown by \citet{Kwok23b}, these features are well-explained by two Gaussian components of [\ion{Fe}{2}] offset by $\sim 5{,}000$~\kms and with FWHM of $\sim 5{,}500$~\kms. The peaks of these IGE emission profiles are redshifted, while the peaks of the IME emission profiles are blueshifted. This could be the result of an asymmetric explosion, like a violent merger, that produced different compositions of ejecta moving in opposite directions along our line of sight.

\subsection{Early-Phase CSM Interaction}

\begin{figure}
    \centering
    \includegraphics[width=3.2in]{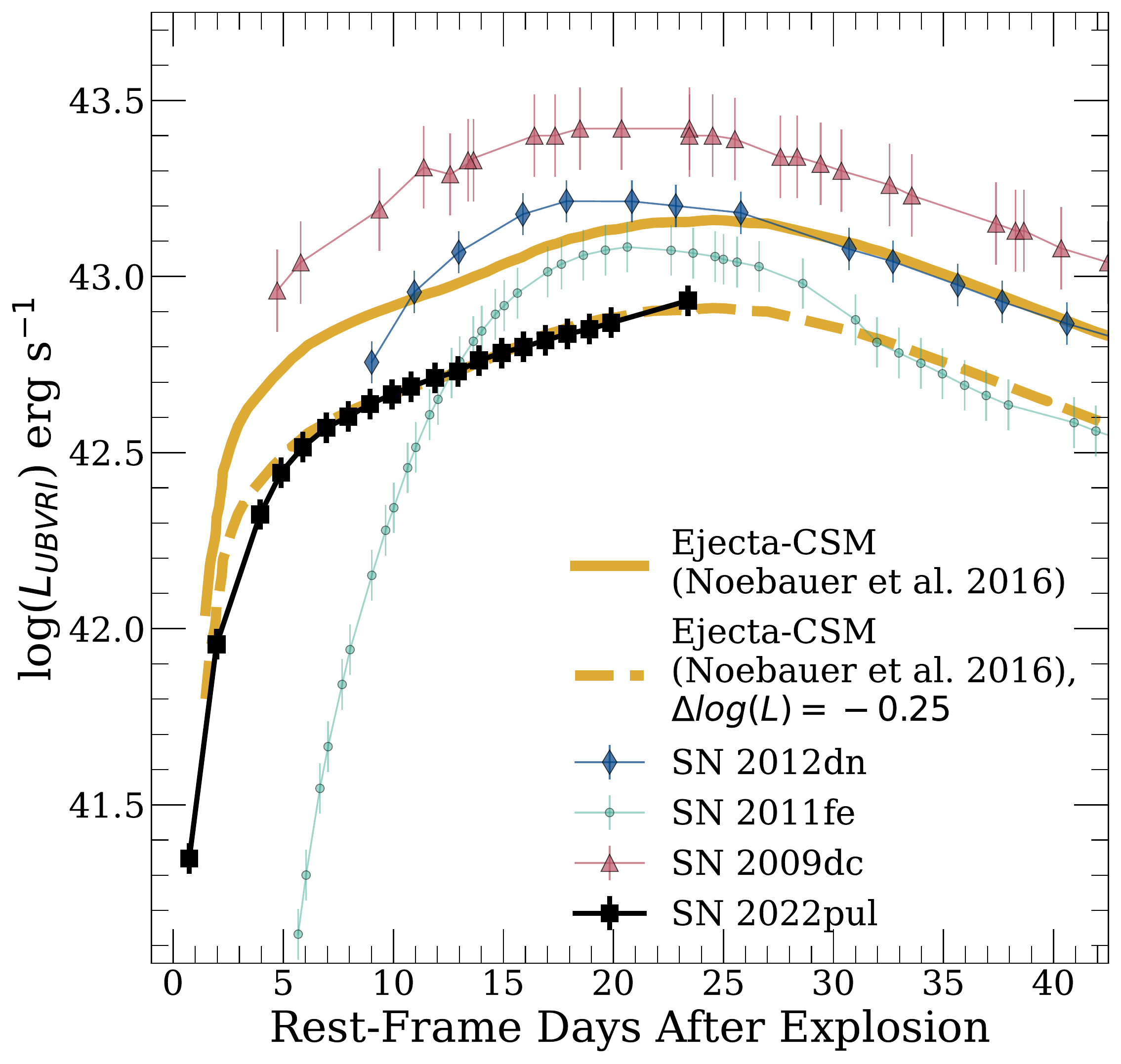}
    \caption{The $UBVRI$ pseudobolometric light curve of \pul\ (black squares) compared with the ejecta-CSM model from \citet{Noebauer16} (yellow curve), the relatively low-luminosity 03fg-like SN~2012dn (blue diamonds), the high-luminosity 03fg-like SN~2009dc (red triangles), and SN~2011fe (green circles). We have shifted the ejecta-CSM model by 0.25 dex (dashed yellow curve) to highlight the good agreement of the light-curve shape of \pul\ with the model.
    \label{fig:ejecta_csm}}
\end{figure}

We examine the early-phase photometry of \pul\ in more detail in \autoref{fig:ejecta_csm}. To compare with models, we constructed an estimated $UBVRI$ pseudobolometric light curve (black squares) from the observed photometry (see \autoref{fig:lightcurve}). We adopted a spectral energy distribution (SED) from the early-time spectra, adjusted it to match the photometry at each epoch, and integrated in the range 3000--9000~\AA\ to estimate the pseudobolometric flux. This is compared with similarly constructed pseudobolometric light curves of SN~2009dc (red triangles), SN~2012dn (blue diamonds), and SN~2011fe (green circles). The solid yellow curve is the synthetic $UBVRI$ light-curve model resulting from a normal SN~Ia embedded in a C/O-rich CSM from \citet{Noebauer16}. We have shifted this model down by 0.25 dex (dashed-yellow curve) to highlight the similarity of its light-curve shape to that of \pul. 

In this model, ejecta colliding with the CSM produce a reverse shock that decelerates and compresses the ejecta. The resulting light curve exhibits two distinct components. The steep rise is caused by the rapid release of radiative energy from the initial CSM-interaction phase, then is followed by the longer timescale additional light-curve evolution from radioactive heating.  Radioactive heating then becomes the dominant source of the luminosity, and the light-curve shape in this phase is determined by the amount of $^{56}$Ni produced in the explosion. \citet{Noebauer16} found that the light curve of SN~2009dc could be well-fit with this model by adjusting the composition of the CSM. This provided a luminosity boost that more closely resembled the peak of the light curve of SN~2009dc. Additionally, \citet{Noebauer16} suggested that the low \ion{Si}{2} velocities observed in SN~2009dc relative to normal SNe~Ia could be the result of the deceleration of the ejecta by the CSM, and subsequent conversion of this kinetic energy to radiative energy. 

We find that the ejecta-CSM model from \citet{Noebauer16} is a remarkably good fit to the pseudobolometric light curve of \pul. Similar to the model, \pul\ exhibits a fast initial rise, which then transitions to a shallower slope at $\sim 5$ days after explosion. This clear two-component rise could indicate the CSM is more C/O-rich than that of SN~2009dc \citep{Noebauer16}. \pul\ takes roughly 5 days longer to reach its peak luminosity than SN~2011fe, and its peak luminosity is significantly lower than that of both SN~2012dn and even SN~2011fe. 

Despite the similarity of the light curve to the ejecta-CSM model, it is important to note that we do not see signs of interaction in our early-time optical spectra (\autoref{fig:spec_early}). For example, in the peculiar subclass of SNe~Ia-CSM \citep{Silverman13:csm}, narrow $H\alpha$ emission is present at early times that has been attributed to the presence of H-rich CSM from a nondegenerate companion star. \pul\ only exhibits narrow emission from [\ion{O}{1}] at late times, which is consistent with forming deep within the SN ejecta. 

\section{Discussion}\label{s:disc}

In this section, we first discuss our interpretation of the progenitor scenario of \pul\ from ground-based data alone. In particular, a viable scenario needs to be able to explain the slowly-evolving early-time light curve relative to normal SNe~Ia, the strong \ion{C}{2} absorption in the early spectra, the low ionization state in the nebular phase, and the strong/asymmetric nebular emission from [\ion{O}{1}] and [\ion{Ca}{2}]. 
We then discuss how this evidence is further supplemented by nebular observations in the NIR and MIR from \textit{JWST}, and we briefly discuss our complete optical+NIR+MIR SED, highlighting the main results that are discussed in further detail in Paper II \citep{Kwok23b} and Paper III \citep{Johansson23}. 

\subsection{Ground-Based Evidence for a WD Merger}

Among recent studies of 03fg-like SNe there is a growing evidence that these events can be explained by WD mergers \citep{Dimitriadis22,Srivastav23,Dimitriadis23,Siebert23a}. We summarize some key observational trends in these SNe:
\begin{itemize}
    \item High peak luminosities and broad light curves \citep{Howell06}.
    \item Rapidly evolving early light-curve bumps \citep{Jiang+21_SN20hvf, Dimitriadis23,Srivastav23}.
    \item Faster fading of the optical light curve coincident with excess luminosity in the NIR a few months after peak brightness \citep{Taubenberger13:sc,Taubenberger19, Yamanaka16, Nagao17,Nagao18, Dimitriadis23}.
    \item Strong early \ion{C}{2} absorption and low photospheric velocities relative to normal SNe~Ia \citep{Howell06, Ashall21}. 
    \item Low-ionization-state nebular spectra, including [\ion{O}{1}] emission in some cases \citep{Maeda09,Taubenberger17,Dimitriadis23}.
    \item Asymmetric and blueshifted nebular emission lines (see \citealt{Taubenberger17, Siebert23a}, and Paper II: \citealt{Kwok23b}). 
\end{itemize}

The origin of the term ``super-Chandrasekhar" in reference to 03fg-like SNe was primarily driven by the inferred high luminosity and broad light curve of SN~2003fg \citep{Howell06}. It was proposed that a super-Chandrasekhar-mass WD, supported via rapid rotation, could be formed via accretion in an SD progenitor system \citep{Yoon05}, or via the merger of two WDs \citep{Tutukov94, Howell01}. Simulated observations of the explosions of rapidly-rotating super-Chandrasekhar-mass WDs have a hard time explaining the low velocities seen in these SNe at peak brightness, and do not show the characteristic strong carbon lines of this subclass \citep{Hachinger12, Fink18}.

A total ejecta mass greater than $M_{\rm Ch}$ can also be achieved within a DD progenitor system. DD progenitors have several advantages when it comes to explaining 03fg-like SNe. (1) The limit on the total ejecta mass allows for a variety of double C/O WD systems. (2) The merger of two WDs can be a naturally asymmetric scenario. Observations should be expected to vary significantly along different lines of sight and for systems with different mass ratios. (3) In a violent merger, oxygen is expected to be present in the innermost regions of the ejecta \citep{Kwok23b}. (4) This progenitor scenario offers a potential path to forming C/O-rich CSM prior to explosion via the disruption of the secondary WD. 

In a WD merger scenario, the explosion can occur either pre- or post-merger. In the former case, the explosion occurs during  the dynamical interaction phase of the two WDs (alternatively referred to as  ``prompt-mergers," ``perimergers," or ``violent mergers" in the literature; \citealt{Pakmor10, Pakmor12, Moll14}); in the latter case, the explosion occurs after the disruption of the secondary WD \citep{Raskin13}.  Hydrodynamical models of post-mergers can produce a significant amount of unburned carbon and oxygen as a result of tidal stripping of the secondary WD \citep{Raskin13}. In either scenario, observations are expected to vary greatly with viewing angle. \citet{Moll14} found that for prompt mergers, depending on viewing angle and the progenitor mass ratio, peak luminosities can vary by a factor of two and span most of the luminosity range of 03fg-like SNe. Furthermore, in post-merger simulations, \citet{Raskin14} showed that the distribution of IMEs and IGEs in the ejecta is ``hourglass"-shaped and surrounded by a dense, C/O-rich CSM. A significant fraction of oxygen is also present in the inner regions of the ejecta.

Early-time observations of some 03fg-like SNe show fast-evolving light-curve bumps that have been interpreted as interaction with a dense CSM \citep{Jiang+21_SN20hvf, Srivastav23, Dimitriadis23}.  \citet{Dimitriadis22} found that the model light curve of their higher mass model ($1.2\ M_{\odot}$ + $1.0\ M_{\odot}$), when viewed along the equator, matched the peak luminosity of the 03fg-like SN~2020esm, but not the shape of its light curve. These models did not consider interaction of the ejecta with the CSM, which can provide additional luminosity and modulate the light-curve shape \citep{Noebauer16}.

The ground-based observations of \pul\ add to the growing evidence that the progenitors of 03fg-like SNe come from WD mergers occurring within a dense C/O-rich CSM, while adding significantly to the diversity of this subclass. Specifically, in both the photospheric and nebular phases, SN~2022pul is most similar to the 03fg-like SN~2012dn with some key differences. In the nebular phase, \pul\ displays strong, broad [\ion{Ca}{2}] emission that is completely blended with [\ion{Fe}{2}] and [\ion{Ni}{2}], unlike any other 03fg-like SN~Ia. 

The primary Ca component of this feature is well-explained by a single Gaussian component and offset by about $-2{,}000$~\kms\ from the systemic velocity of the host galaxy. Symmetric components of [\ion{Ca}{2}] have been observed only in ASASSN-15pz \citep{Chen19, Siebert23a}, SN~2020esm \citep{Dimitriadis22}, and SN~2021zny \citep{Dimitriadis23}. \citet{Siebert23a} argued that the multiple broad velocity components of [\ion{Ca}{2}] observed in SN~2020hvf were produced by two distinct ejecta components offset by the orbital velocity of the WDs prior to explosion. This type of emission is more likely to be observed if the line of sight was along the orbital plane. The symmetric component of [\ion{Ca}{2}] could indicate that we are instead observing \pul\ along the pole, where you would not expect to be capable of seeing velocity offsets. However, this interpretation is inconsistent with the velocity offsets observed between NIR and MIR emission profiles from IMEs and IGEs \citep{Kwok23b}. Additionally, all clearly identified lines from IGEs in the optical and NIR nebular spectra are asymmetric relative to what is seen in normal SNe~Ia, further supporting an asymmetric explosion scenario. 

One of the most striking features in the ground-based nebular spectra is the strong [\ion{O}{1}] emission, which is extremely rarely observed in thermonuclear SNe. Several sub-Chandrasekhar-mass explosion models predict strong [\ion{Ca}{2}] and low ionization states (see \citealt{Wilk20}, discussed further in Paper II: \citealt{Kwok23b}, and \citealt{Polin21:neb}), but a merger is the only model that predicts [\ion{O}{1}] at low velocities as seen in \pul. Similar emission has been observed in low-luminosity subclasses such as 02es-like SNe~Ia and Ca-rich SNe. Specifically for the 02es-like SNe~Ia, both \citet{Taubenberger13:10lp} and \citet{Kromer16} favor a violent merger model for SN~2010lp and iPTF14atg (respectively), as discussed in Section \ref{sec:OI}. The significant amount of [\ion{O}{1}] present at low velocity is best explained by this scenario. 

The evolution of the optical nebular spectra displayed in \autoref{fig:spec_time} shows a clear progression from an SED that appears 03fg-like to an SED dominated by [\ion{O}{1}] and [\ion{Ca}{2}]. Fast changes in ionization state, including the appearance of broad [\ion{Ca}{2}], have been observed in normal SNe~Ia at very late times, which has been attributed to clumping of the ejecta \citep{Tucker22a}. However, our latest phase optical spectrum of \pul\ is quite extreme in comparison, bearing a closer resemblance to Ca-rich SNe. For \pul, we investigate if additional clumping of the ejecta can improve the match of violent merger models to the data in Paper II \citep{Kwok23b}. Double-degenerate progenitor scenarios have also been favored for some Ca-rich SNe, though not violent mergers specifically \citep{Jacobson-Galan19,Zenati+23_carich}. Despite the low luminosities of these peculiar subclasses, the similarity of their nebular emission to that of \pul\ is intriguing and warrants further study.

If we combine our evidence for a violent merger with the unique early light curve of \pul\ we can begin to build a picture of its progenitor. \pul\ was likely the result of the merger of two C/O WDs within dense CSM. This result is corroborated by nebular-phase MIR observations of \pul\ with \textit{JWST}.

\subsection{Continuum Emission Observed with \textit{JWST}}

In \autoref{fig:full_spec}, we presented our complete optical+NIR+MIR nebular-phase spectrum at 338 days after explosion. With MIRI, we observe a strong blackbody continuum ($T \approx 500 K$; \citealt{Johansson23}). 
The existence of  C/O-rich CSM may provide the necessary conditions for dust formation, and several studies have suggested that this can explain some of the important trends in ground-based observations of 03fg-like SNe \citep{Taubenberger13:sc,Taubenberger19,Hsiao20,Ashall21,maeda23}. In particular, an accelerated decline in the optical light curves, potentially due to the onset of dust formation in the SN ejecta, has been observed beginning as early as $\sim 60$ days after peak brightness \citep{Taubenberger13:sc,Dimitriadis22}. \citet{Hsiao20} suggested that the timing of the onset of this effect could be related to the mass of the C/O-rich envelope. Additionally, an excess in NIR light curves has been observed, which might support this theory of dust formation while the pre-existing CSM dust echo is another possibility \citep{Yamanaka16,Nagao16}. Unfortunately, we are unable to constrain these light-curve effects with the available data of \pul. More work is needed to understand if the optical line shapes in the nebular phase are being affected by dust extinction. For a detailed discussion of the dust emission observed in \pul\ see Paper III \citep{Johansson23}. 

One alternative explanation for the dense CSM distribution and subsequent dust formation involves the explosion of a Chandrasekhar-mass WD within the envelope of an AGB star (the core-degenerate scenario; see \citealt{Kashi_Soker11,Hsiao20,Ashall21}). This model predicts a large X-ray luminosity caused by interaction with the envelope that has not yet been observed. Furthermore, due to the high-density burning that occurs in Chandrasekhar-mass explosions \citep{Hoflich96}, these SNe may be expected to form more stable $^{58}$Ni. Nebular emission lines of nickel are easier to detect in the MIR, and have been observed in \textit{JWST} spectra of a normal SN~Ia (SN~2021aefx; \citealt{Kwok+23K,DerKacy+23}). Emission from stable $^{58}$Ni in \pul\ is quite weak relative to the classical Type Ia SN~2021aefx, further supporting the idea that \pul\ was produced by the merger of lower-mass C/O WDs. For a complete discussion of the nebular-line ionization states, asymmetries, and comparison to explosion models see Paper II \citep{Kwok23b}.

Finally, while dust formation is plausible \citep{Johansson23}, we do not see changing attenuation of the redshifted material in the nebular optical emission lines of \pul, which has been observed in dust-forming SNe like SN~2006jc \citep{Smith08:06jc}. A MIR excess has also been observed in Type Iax supernova, SN~2014dt \citep{Fox16}. \citet{Foley16:iax} found that the nebular properties of SNe~Iax are consistent with the presence of a bound remnant, and that the MIR properties of SN~2014dt were better explained by this model. \citet{Siebert23a} suggested that the extremely narrow nebular [\ion{Ca}{2}] (similar to what has been observed in SNe~Iax) emission observed in the 03fg-like SN~2020hvf could be explained by a wind from a bound remnant. This possibility should continue to be considered in when analyzing future observations of 03fg-like SNe. 

\section{Conclusions}\label{s:conc}

The specific progenitor systems of thermonuclear SNe have remained elusive for decades. Currently, there is still debate over the types of systems and explosion mechanisms that contribute to the majority of the population of normal SNe~Ia. 

Our thorough ground- and space-based follow-up campaign of \pul\ has provided some strong constraints on an SN~Ia progenitor system and greatly informs our understanding of the carbon-strong, high-luminosity 03fg-like subclass of SNe~Ia. Specifically, the multicomponent rise in the optical light curve, narrow nebular [\ion{O}{1}] emission at low velocity, low ionization state and asymmetric nebular features, weak emission from stable $^{58}$Ni \citep{Kwok23b}, and strong emission from dust \citep{Johansson23}, are all consistent with the merger of two WDs within C/O-rich CSM. 

The diversity of 03fg-like SNe~Ia is also consistent with this naturally asymmetric scenario. Observations of these explosions should be expected to vary significantly in their luminosity and velocity distributions along different lines of sight. Additionally, the total mass of the system will affect the luminosity, and the mass ratio could lead to different dust distributions and CSM masses. If the conditions of the merger affect the amount of mass from the secondary WD that ends up in the CSM versus the central region of the ejecta, this might explain the varying amounts of [\ion{O}{1}] emission that have been observed in the nebular spectra of 03fg-like SNe. 

The detection of MIR excesses in now three 03fg-like SNe~Ia (SN~2009dc, ASASSN-15pz, and \pul; see Paper III: \citealt{Johansson23}) shows that \textit{JWST} is an important resource for their follow-up. Future observations should aim to obtain MIR spectral sequences of these SNe to constrain the onset of dust formation, and whether it was formed in the SN ejecta or perhaps in pre-existing CSM. Furthermore, the nebular emission features in the MIR are much less sensitive to temperature structure than optical lines, and provide critical information on the ionization and elemental distributions in the ejecta \citep{DerKacy+23, Kwok+23K, Kwok23b}. Higher-resolution observations are needed to probe the presence of narrow emission features, an indication that some of these SNe may have bound remnants \citep{Siebert23a}. 

In addition to late-time MIR observations, extremely early UV and optical observations of 03fg-like SNe are needed to adequately understand the CSM interaction phase which can only last a few days \citep{Jiang+21_SN20hvf, Srivastav23,Dimitriadis23}. Prolonged interaction with an extended envelope (i.e., in the core-degenerate scenario) is expected to produce a large X-ray luminosity \citep{Hsiao20}. Detailed modeling of this scenario is needed, especially in the nebular phase, to understand if it can reproduce the observed properties of 03fg-like SNe~Ia. Additionally, currently 03fg-like SNe do not show strong continuum polarization \citep{Bulla16}; polarimetry at earlier times is essential to understanding the asymmetry of the explosion. 

Finally, one large caveat regarding the merger scenario is the amount of CSM that can be formed prior to explosion. The ejecta-CSM model \citep{Noebauer16} discussed in \autoref{s:anal} used a CSM mass of 0.6~$M_{\odot}$. It may be difficult to reconcile this large mass with the substantial amount of oxygen in the central region of the ejecta and the observed luminosity range of 03fg-like SNe. Moreover, the dynamical double WD models of \citet{Raskin13} had less than $5\times10^{-3}\,M_{\odot}$ of CSM, likely insufficient to produce the interaction signature seen in \pul. Further modeling is needed to explore the parameter space of WD mergers to find a scenario well-matched to the observations of this diverse class of supernova explosions. \\
\\
\\

This work is based on observations made with the NASA/ESA/CSA \textit{JWST} as part of program \#02072. We thank Shelly Meyett for her consistently excellent work scheduling the \textit{JWST} observations, Sarah Kendrew for assistance with the MIRI observations, and Glenn Wahlgren for help with the NIRSpec observations. 

The data were obtained from the Mikulski Archive for Space Telescopes at the Space Telescope Science Institute (STScI), which is operated by the Association of Universities for Research in Astronomy (AURA), Inc., under National Aeronautics and Space Administration (NASA) contract NAS 5-03127 for \textit{JWST}. Support for this program at Rutgers University was provided by NASA through grant JWST-GO-02072.001. 

The SALT data presented herein were obtained with Rutgers University program 2022-1-MLT-004 (PI S.~W.~Jha). We are grateful to SALT Astronomer Rosalind Skelton for taking these observations.

This work makes use of data from the Las Cumbres Observatory global network of telescopes.  The LCO group is supported by NSF grants AST-1911151 and AST-1911225. This work also makes use of data gathered with the 6.5 meter Magellan telescopes at Las Campanas Observatory, Chile.

M.R.S. is supported by an STScI Postdoctoral Fellowship. G.D. acknowledges H2020 European Research Council grant \#758638. L.A.K. acknowledges support by NASA FINESST fellowship 80NSSC22K1599. C.L. is supported by an NSF Graduate Research Fellowship under grant \# DGE-2233066.

The UCSC team is supported in part by NASA grant NNG-17PX03C, National Science Foundation (NSF) grant AST-1815935, the Gordon and Betty Moore Foundation, the Heising-Simons Foundation, and a fellowship from the David and Lucile Packard Foundation to R.J.F. The work of A.V.F.'s supernova group at UC Berkeley is generously supported by the Christopher R. Redlich Fund, Alan Eustace (W.Z. is a Eustace Specialist in Astronomy), Frank and Kathleen Wood (T.G.B. is a Wood Specialist in Astronomy), Gary and Cynthia Bengier, Clark and Sharon Winslow, Sanford Robertson (Y.Y. is a Bengier-Winslow-Robertson Postdoctoral Fellow), and many other donors.  

S.B. acknowledges support from the Alexander von Humboldt Foundation and from the ``Programme National de Physique Stellaire'' (PNPS) of CNRS/INSU cofunded by CEA and CNES.
D.J.H. receives support through NASA astrophysical theory grant 80NSSC20K0524.
J.V. and T.S. are supported by the NKFIH-OTKA grants K-142534 and FK-134432 of the Hungarian National Research, Development and Innovation (NRDI) Office, respectively.
B.B. and T.S. are supported by the \'UNKP-22-4 and \'UNKP-22-5 New National Excellence Programs of the Ministry for Culture and Innovation from the source of the NRDI Fund, Hungary. T.S. is also supported by the J\'anos Bolyai Research Scholarship of the Hungarian Academy of Sciences.
The research of J.C.W. and J.V. is supported by NSF grant AST-1813825.

A.F. acknowledges support by the European Research Council (ERC) under the European Union’s Horizon 2020 research and innovation program (ERC Advanced Grant KILONOVA \#885281).
M.D., K.M., and J.H.T. acknowledge support from EU H2020 ERC grant \#758638.
Research by Y.D. and S.V. is supported by NSF grant AST-2008108

Time-domain research by D.J.S. and the University of Arizona team is supported by NSF grants AST-1821987, 1813466, 1908972, and 2108032, and by the Heising-Simons Foundation under grant \#2020--1864. 

K.M. acknowledges support from the Japan Society for the Promotion of Science (JSPS) KAKENHI grant JP20H00174, and the JSPS Open Partnership Bilateral Joint Research Project (JPJSBP120209937). 

L.G. acknowledges financial support from the Spanish Ministerio de Ciencia e Innovaci\'on (MCIN), the Agencia Estatal de Investigaci\'on (AEI) 10.13039/501100011033, and the European Social Fund (ESF) ``Investing in your future'' under the 2019 Ram\'on y Cajal program RYC2019-027683-I and the PID2020-115253GA-I00 HOSTFLOWS project, from Centro Superior de Investigaciones Cient\'ificas (CSIC) under the PIE project 20215AT016, and the program Unidad de Excelencia Mar\'ia de Maeztu CEX2020-001058-M.

J.P.H. acknowledges support from the George A.\ and Margaret M.\ Downsbrough bequest. The Aarhus supernova group is funded in part by an Experiment grant (\#28021) from the Villum FONDEN, and by a project 1 grant (\#8021-00170B) from the Independent Research Fund Denmark (IRFD).

This publication was made possible through the support of an LSSTC Catalyst Fellowship to K.A.B., funded through grant 62192 from the John Templeton Foundation to the LSST Corporation. The opinions expressed in this publication are those of the authors and do not necessarily reflect the views of LSSTC or the John Templeton Foundation.

Some of the data presented herein were obtained at the W.~M. Keck Observatory, which is operated as a scientific partnership among the California Institute of Technology, the University of California, and NASA. The Observatory was made possible by the generous financial support of the W.~M. Keck Foundation. The authors wish to recognize and acknowledge the very significant cultural role and reverence that the summit of Maunakea has always had within the indigenous Hawaiian community. We are most fortunate to have the opportunity to conduct observations from this mountain.

A major upgrade of the Kast spectrograph on the Shane 3~m telescope at Lick Observatory was made possible through generous gifts from the Heising-Simons Foundation as well as William and Marina Kast. Research at Lick Observatory is partially supported by a generous gift from Google.

We thank the Subaru staff for the data taken by the Subaru Telescope (S23A-023). 
% \end{acknowledgments}  

\facilities{AAVSO, ANU (WiFeS), ASAS-SN, ATLAS, GTC (OSIRIS), JWST (NIRSpec/MIRI), Keck:I (LRIS), Keck:II (NIRES), Keck:II (DEIMOS), LCO/GSP, Magellan (IMACS), MMT (Binospec), SALT (RSS), Shane (Kast), SOAR (Goodman), Subaru (FOCAS), UH (SNIFS), ZTF}
\software{astropy \citep{astropy,astropy2,astropy3}, {\tt YSE-PZ} \citep{CoulterZenodo, CoulterYSEPZ}}
\newpage
\appendix \label{app}
\begin{table*}[htb!]
\centering
\begin{tabular}{ccccc}
\hline
 Source   &        MJD & Filter   &   Brightness (mag) &   Uncertainty (Mag) \\
\hline
 ASAS-SN  & 59786.7343 & g        &              15.71 &                0.06 \\
 ASAS-SN  & 59786.7346 & g        &              15.88 &                0.06 \\
 ASAS-SN  & 59786.7346 & g        &              15.83 &                0.06 \\
 ASAS-SN  & 59787.9733 & g        &              14.48 &                0.02 \\
 ASAS-SN  & 59787.9737 & g        &              14.50 &                0.02 \\
 ASAS-SN  & 59787.9737 & g        &              14.51 &                0.02 \\
 ASAS-SN  & 59789.9559 & g        &              13.51 &                0.02 \\
 ASAS-SN  & 59790.7063 & g        &              13.16 &                0.01 \\
 ASAS-SN  & 59790.7066 & g        &              13.23 &                0.01 \\
 ASAS-SN  & 59790.7066 & g        &              13.21 &                0.01 \\
 AAVSO    & 59790.9677 & I        &              13.17 &                0.06 \\
 ASAS-SN  & 59790.9682 & g        &              13.22 &                0.01 \\
 AAVSO    & 59790.9683 & I        &              13.09 &                0.05 \\
 ASAS-SN  & 59790.9685 & g        &              13.17 &                0.01 \\
 ASAS-SN  & 59790.9685 & g        &              13.16 &                0.01 \\
 AAVSO    & 59790.9690 & V        &              13.07 &                0.01 \\
 AAVSO    & 59790.9697 & V        &              13.05 &                0.01 \\
 AAVSO    & 59790.9706 & B        &              13.10 &                0.02 \\
 AAVSO    & 59790.9717 & B        &              13.08 &                0.02 \\
 AAVSO    & 59790.9726 & R        &              13.04 &                0.02 \\
 AAVSO    & 59790.9731 & R        &              13.04 &                0.02 \\
\hline
\end{tabular}
\caption{Log of ground-based photometry of SN~2022pul. This table is available in its entirety in machine-readable form.}
\label{tab:1}
\end{table*}
\begin{table}[htb!]
\centering
\begin{tabular}{cccrc}
\hline
 Telescope   & Instrument   &     MJD &   Phase & Wavelength Range   \\
\hline
 ANU 2.3m      & WiFeS        & 59790.4 &     -17 & 3789 - 8972        \\
 UH 2.2m       & SNIFS        & 59792.3 &     -15 & 4054 - 9073        \\
 Lick        & Kast         & 59793.2 &     -14 & 3247 - 10862       \\
 Keck II      & NIRES        & 59944.6 &    +137 & 9383 - 24615       \\
 Lick        & Kast         & 59975.5 &    +168 & 3244 - 10862       \\
 Keck II      & ESI          & 59990.0 &    +182 & 3938 - 10168       \\
 Keck II      & ESI          & 60030.0 &    +222 & 3938 - 10168       \\
 Keck II      & NIRES        & 60033.4 &    +226 & 9383 - 24613       \\
 Lick        & Kast         & 60046.4 &    +239 & 3244 - 10861       \\
 Subaru     & FOCAS        & 60048.0 &    +240 & 3642 - 9968        \\
 Lick        & Kast         & 60055.4 &    +248 & 3609 - 10706       \\
 Lick        & Kast         & 60057.4 &    +250 & 3244 - 10862       \\
 Keck II      & DEIMOS       & 60057.5 &    +267 & 4386 - 9597        \\
 Lick        & Kast         & 60062.4 &    +255 & 3611 - 10706       \\
 Keck II      & DEIMOS       & 60075.3 &    +267 & 4753 - 7393        \\
 Lick        & Kast         & 60083.4 &    +276 & 3493 - 10861       \\
 Magellan    & IMACS        & 60078.0 &    +270 & 4217 - 9388        \\
 SOAR        & Goodman      & 60099.0 &    +291 & 4922 - 8909        \\
 Keck II      & NIRES        & 60102.4 &    +294 & 9383 - 24615       \\
 SOAR        & Goodman      & 60112.1 &    +304 & 3990 - 6979        \\
 SALT        & RSS          & 60117.0 &    +309 & 3488 - 7237        \\
 Lick        & Kast         & 60117.2 &    +309 & 3611 - 10702       \\
 MMT         & Binospec     & 60118.0 &    +310 & 3815 - 9185        \\
 Magellan    & IMACS        & 60120.0 &    +312 & 3789 - 9451        \\
 JWST        & NIRspec      & 60124.0 &    +316 & 6018 - 49967      \\
 JWST        & MIRI         & 60124.0 &    +316 & 49645 - 140457      \\
 GTC         & OSIRIS       & 60126.9 &    +319 & 3627 - 10261       \\
 Lick        & Kast         & 60144.2 &    +336 & 3625 - 10688       \\
\hline
\end{tabular}
\caption{Log of spectroscopic observations of \pul.}
\label{tab:spec}
\end{table}

\newpage
\bibliography{astro_refs}
\bibliographystyle{aasjournal}

\end{document}